\documentclass[sigconf]{acmart}

\usepackage{booktabs} 
\usepackage{verbatim}
\usepackage{graphicx}
\usepackage{multirow}
\usepackage{array}
\usepackage{subfigure}
\usepackage{multicol}
\usepackage{color}
\usepackage{epsfig}
\usepackage{natbib}
\usepackage{xspace}
\usepackage{booktabs}
\usepackage{url}

\copyrightyear{2018}
\acmYear{2018} 
\setcopyright{iw3c2w3}
\acmConference[WWW 2018]{The 2018 Web Conference}{April 23--27, 2018}{Lyon, France}
\acmBooktitle{WWW 2018: The 2018 Web Conference, April 23--27, 2018, Lyon, France}
\acmPrice{}
\acmDOI{10.1145/3178876.3186144}
\acmISBN{978-1-4503-5639-8/18/04}

\newcommand{\hide}[1]{} 
\newcommand{\vpara}[1]{\vspace{0.01in}\noindent\textbf{#1 }}
\newcommand{\para}[1]{\vspace{0.01in}\noindent\textbf{#1 }}
\newcommand{\secref}[1]{Section~\ref{#1}} 
\newcommand{\figref}[1]{Figure~\ref{#1}} 

\newcommand{\local}{local\xspace}

\newcommand{\locals}{locals\xspace}
\newcommand{\stays}{staying migrants\xspace}
\newcommand{\leaves}{leaving migrants\xspace}

\newcommand{\Locals}{Locals\xspace}
\newcommand{\Stays}{Staying migrants\xspace}
\newcommand{\Leaves}{Leaving migrants\xspace}

\begin{document}

\title{To Stay or to Leave: Churn Prediction for \\ Urban Migrants in the Initial Period}

\author{Yang Yang}
\affiliation{
  \institution{Zhejiang University}
}
\email{yangya@zju.edu.cn}

\author{Zongtao Liu}
\affiliation{
  \institution{Zhejiang University}
}
\email{tomstream@zju.edu.cn}

\author{Chenhao Tan}
\affiliation{
  \institution{University of Colorado Boulder}
}
\email{chenhao@chenhaot.com}

\author{Fei Wu}
\affiliation{
  \institution{Zhejiang University}
}
\email{wufei@zju.edu.cn}

\author{Yueting Zhuang}
\affiliation{
  \institution{Zhejiang University}
}
\email{yzhuang@zju.edu.cn}

\author{Yafeng Li}
\affiliation{
  \institution{China Telecom.}
}
\email{liyafeng@chinatelecom.cn}

\renewcommand{\shorttitle}{To Stay or to Leave: Churn Prediction for Urban Migrants in the Initial Period}
\renewcommand{\shortauthors}{Y. Yang et al.}

\setlength{\floatsep}{6pt}


\begin{abstract}
In China, 260 million people migrate to cities to realize their urban dreams. 
Despite 
that these migrants
play an important role in the rapid urbanization process, many of them fail to settle down and eventually leave the city. 
The integration process of migrants thus raises an important issue 
for scholars and policymakers. 

In this paper, we use Shanghai as an example to investigate migrants' behavior in their first weeks and in particular, how their behavior relates to early departure. 
Our dataset consists of 
a one-month complete dataset of 698 telecommunication logs
between 54 million users,
plus a novel and publicly available housing price data for 18K real estates in Shanghai.
We find that 
migrants who end up leaving early tend to neither develop diverse connections in their first weeks nor move around the city.
Their active areas also have higher housing prices than that of 
\stays.
We formulate a churn prediction problem to determine 
whether a migrant is going to leave based on her behavior in the first few days.
The prediction performance improves as we include data from more days.
Interestingly, when using the same features, the classifier trained from only the first few days is already as good as the classifier trained using full data, suggesting that the performance difference mainly lies in the difference between features.
\end{abstract}



\maketitle


\section{Introduction}
\label{sec:intro}

\smallskip
\small
\hfill {\em In a big city like L.A. you can spend a lot of time surrounded by hundreds of people yet you feel like an alien or a ghost or something.}

\hfill --- {\em Morley}
\normalsize
\bigskip

Millions of people migrate to cities to realize their urban dreams, ranging from pursuing potential job opportunities to embracing an open dynamic culture \cite{lee2015world}.
These migrants 
contribute to the prosperity of cities by
constituting a substantial part of the workforce in the cities,
strengthening the political and economic status of the cities,
and bringing diverse cultures to the cities.

Despite the great benefits brought by migration, policymakers and scholars have well recognized that the fast rate of migration poses great challenges \cite{Bai:NatureNews:2014,lee2015world}. 
Segregation and social inequality have become significant issues in the migration process.
For instance, 
migrants may settle in slums with health hazards \cite{beguy2010circular};
they tend to be overworked but underpaid \cite{razavi2010underpaid};
their children may be excluded from schools \cite{Goodburn:InternationalJournalOfEducationalDevelopment:2009}.
These problems might be even more salient in China, a developing country with an unprecedented speed of urbanization \cite{Bai:NatureNews:2014}. 
It is thus an important research question to understand the integration of migrants into urban society.

In this paper, we focus on the initial period of a migrant's integration process because a migrant's first steps are important for her eventual integration.
Despite the importance of the initial period~\cite{beguy2010circular, chandrasekhar2015short-term, newbold2007short-term}, existing studies, which mostly rely on survey data, rarely have fine-grained data to examine this period. 
We take Shanghai as an example to 
investigate two aspects based on telecommunication data:
how they develop their initial personal networks and how they move around the city. 
In particular, our dataset allows us to explore why some migrants decided to leave early.
This problem of whether to stay in a new city resembles studies on whether users will stay in online communities, also known as churn prediction \cite{lampe2005follow,McAuley:2013:ACM:2488388.2488466,Danescu-Niculescu-Mizil:2013:NCO:2488388.2488416,tan2015all,Backstrom:2006:GFL:1150402.1150412,Crandall:2008:FES:1401890.1401914}, but presents more complex dynamics because moving offline requires considerable amount of efforts.

\vpara{Organization and highlights of this work.} 
We present a large-scale quantitative exploration on the first weeks after migrants arrive in Shanghai, one of the biggest cities in China. 
We employ a Shanghai {\em one-month complete} telecommunication metadata provided by China Telecom, the third largest mobile service providers in China. 
Our dataset consists of around 698 million call logs between 54 million users. 
In addition, we collect housing prices of over 20K real estates in Shanghai to study the role of housing price in migrant integration. 
The details of our datasets are introduced in Section~\ref{sec:setup}.
We are able to identity whether a user of China Telecom is a migrant because 1) it is necessary for a migrant to apply for a local phone number in Shanghai due to long-distance costs; 2) it is uncommon for a temporary visitor to apply for a local phone number due to the burdensome application process; 3) applying for phone numbers require personal identification, which contains birthplace information.
We use {\em locals}, who were born in Shanghai, as a comparison point to understand the integration process.
We also differentiate {\em \leaves}, new migrants that left the city in three weeks after they moved to Shanghai, from {\em \stays}, new migrants that managed to stay for the first month.   
Our results indicate that around 4\% of new migrants ended up leaving early.
This work builds on our previous work~\cite{yang2018}, which employs the same telecommunication dataset to explore different characteristics of new migrants, settled migrants, and locals. 
The key differences lie in the churn prediction task to predict the early departure of new migrants and the novel combination of housing price information and telecommunication metadata.

We first explore how \locals, \leaves, and \stays differ in their mobile communication networks and geographical locations in \secref{sec:analysis}.
The dynamic patterns over time allow us to study the integration process in the first month.
Overall, \locals are stable in all our proposed features during this period, whereas both \stays and \leaves go through significant changes despite the short time span.
However, the changes of \stays and \leaves may happen to different extents, sometimes even in different directions.
Specifically, we find that it is important for a new migrant to develop \emph{diverse} social ties in the first few weeks.
For instance, a migrant with contacts from more provinces tends to stay. 
As for geographical locations, \leaves are less active in terms of geographical movements, and move around more expensive housing areas than both \locals and \stays,
whereas \stays move around cheaper areas over time.
This observation suggests that it is important for new migrants to find their own active area in a big city.
\Stays are still different from \locals in the last week of our dataset, which suggests that one month is far from enough for a migrant to integrate into a new city. 

We then study to what extent we can separate the three groups by formulating prediction tasks in Section~\ref{sec:exp}.
Because of the class imbalance, we investigate two prediction tasks:  distinguishing new migrants from \locals and distinguishing \leaves from \stays.
Our proposed features are effective in both tasks and clearly outperform random guessing.
Random forest is the best performing classifier, indicating the importance of non-linearity.
We focus our analysis on the second task as it holds promise for delivering personalized service to new migrants who find the integration process difficult. 
The prediction performance improves as we include features that span more days. 
Such a performance improvement mainly comes from better feature quality as we use data from more days, because 
the classifier trained from only the first few days can perform as well as the classifier trained using full data by using the same set of features when testing. 

Our work aims to understand the (dis)integration of migrants.
This challenging problem necessarily involves efforts from a wide range of disciplines, including anthropology, 
economics, and sociology.
We thus provide an overview of related work 
in \secref{sec:related} and offer some concluding discussions in \secref{sec:conclude}.


\section{Experimental Setup}
\label{sec:setup}
In this paper, our main dataset is the \textit{complete} telecommunication records between mobile users using China Telecom in Shanghai over one month in 2016. 
Before introducing our dataset and experimental setup, we highlight several facts about telecommunication in China.
First, obtaining a local number is the first integration step for a new migrant because of long-distance call costs,
and we are able to differentiate whether a telephone number is a local number in Shanghai or from other regions.
Second, since obtaining a phone number is nontrivial and requires personal identification, it is uncommon for a temporary visitor to obtain a local number. 
Personal identification allows us to extract the birthplace of a person.
Therefore, we can identify people who just obtained a local number but were not from Shanghai originally.
It is worth noting that long-distance call costs have been removed by China Telecom in September 2017, which makes our telecommunication metadata unique and valuable to understanding migrant integration.
Insights from our work can nevertheless be used to analyze telecommunication patterns without personal identification information.


\subsection{How Many Migrants are Leaving in the First Weeks?}
\label{sec:setup:data}

\vpara{Telecommunication dataset.} Our telecommunication dataset is provided by China Telecom, the third largest mobile service provider in China. 
The dataset spans a month from September 1st, 2016 to September 30th, 2016.
It includes over 698 million call logs between around 54 million users. 
For each user, we obtain her demographical information including age, sex, and birthplace by the personal identification binded with the phone number.
Each entry in the call logs contains the caller's number, the callee's number, the starting time, and the ending time.  
In addition, for each call, we have the GPS location of the corresponding telecommunication tower, which is widely used to approximate user locations during the call.  
\hide{
the GPS location of the corresponding telecommunications tower used 
during the call for users of China Telecom, which roughly approximates the locations of them.
}
Our dataset was anonymized by China Telecom to protect user privacy. 
Throughout the paper, we report only average statistics without revealing any identifiable information of individuals.

\begin{figure}[t]
\centering
\epsfig{file=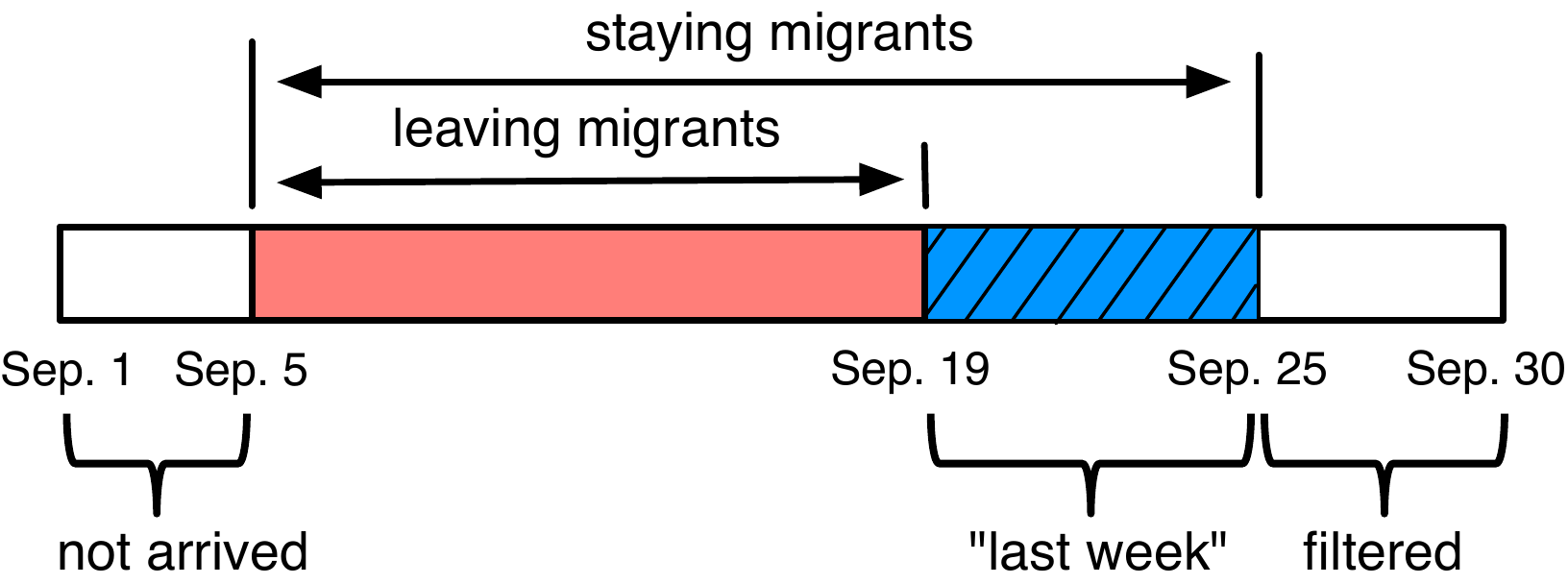, width=0.45\textwidth}
\caption{An illustration of how we define \leaves and \stays. The first several days are used to filter out new migrants. Since the last few days overlap with national holidays, we use Sep. 19 --- Sep. 25 as the last week to make sure that leaving migrants left early instead of traveling temporarily.}
\label{fig:define}
\end{figure} 

\vpara{\Locals, \stays, and \leaves.} 
We only consider users with local phone numbers in this work.\footnote{We merge numbers corresponding to the same user ID into one to account for users with multiple numbers. We also filtered around $15,000$ users that have abnormally high degrees, who likely correspond to fraudsters, delivery persons, or customer services according to a user type list provided by China Telecom.}
We categorize users in our dataset into three groups based on their birthplaces and call history.
We refer to people that were born in Shanghai as \textit{\locals}. 
We consider people that were not born in Shanghai and had no call logs in the first 4 days in our dataset as {\em new migrants}.  
Our focus in this paper is to understand the behavioral pattern of new migrants, which shed light on the integration process of migrants.

Our first question is how many new migrants are leaving in the first weeks, despite that they made efforts to obtain a local number.
We identify new migrants that ended up leaving Shanghai early, i.e., before {\em the last week} in our dataset.
To make sure that people did not leave temporarily, 
we omit the last 5 days' data for all users as 
the National Day holidays were close to that time, which may lead to temporal travel.
That is, the last week in our dataset is defined as Sep. 19 - Sep. 25. 
We consider new migrants as {\em \leaves} if they were active in the first two weeks (Sep. 5 - Sep. 18) and have no record since Sep. 19,
and as {\em \stays}  if they were active in all the three weeks. 
Figure~\ref{fig:define} gives an illustration. 


Based on our definition, we identify {\em 1.8M \locals, 34K \stays, and 1.5K \leaves}.
It follows that around \textbf{\textit{4\%}} of the new migrants left Shanghai in the first few weeks, which is a useful statistic for urban policymakers and complements existing survey based approaches.
To the best of our knowledge, there is no public official report about this statistic.  
This categorization of users into \locals, \stays, and \leaves constitutes the basis for our computational framework.



\begin{figure}[t]
\raggedleft
\epsfig{file=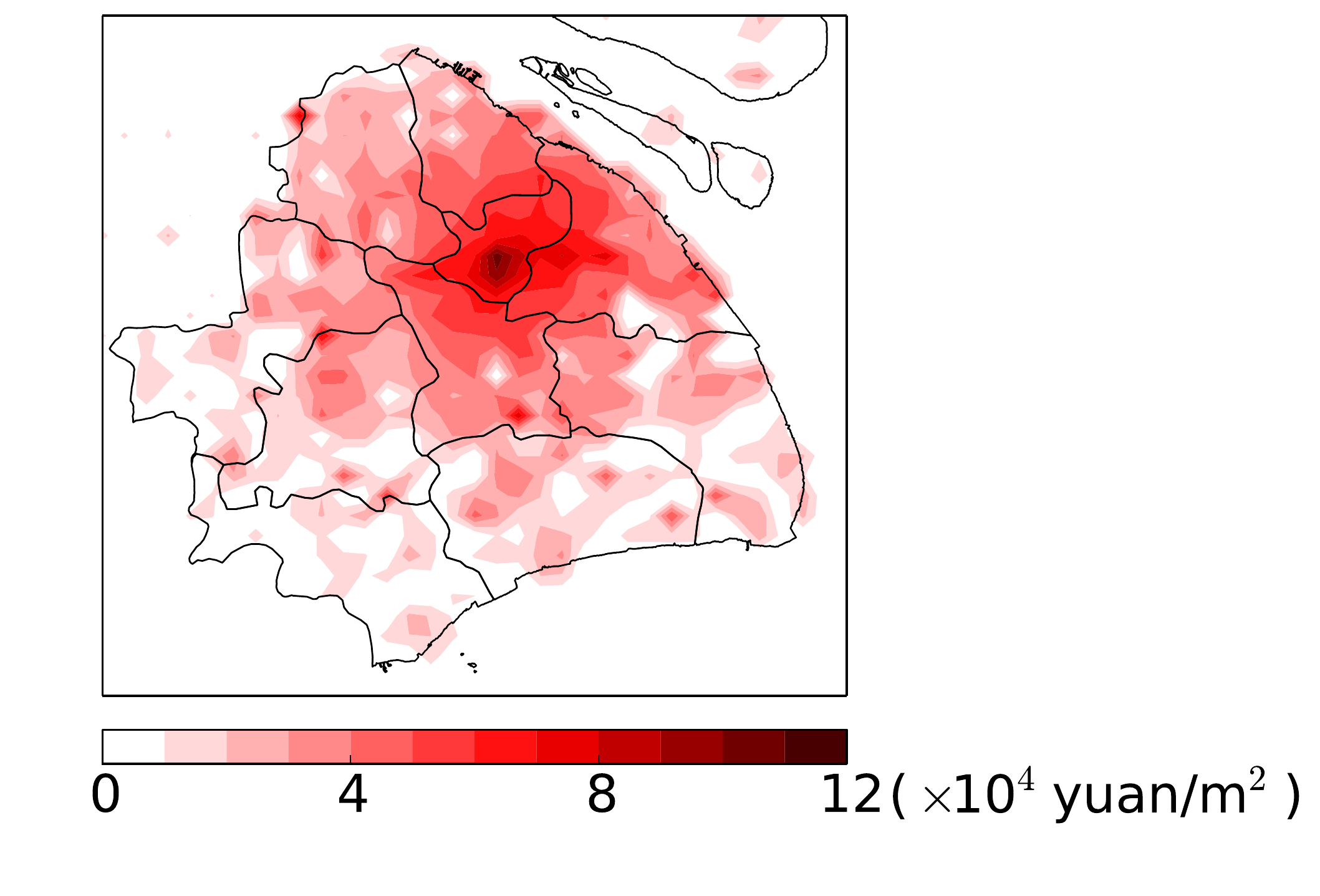, width=0.36\textwidth}
\hspace{0.15in}
\caption{Housing price distribution over Shanghai.}
\label{fig:house}
\end{figure}

\subsection{Housing Price Data} 
Economic theory suggests that individual migration depends on housing prices in different areas~\cite{so2001the}. 
To validate and further study this in our dataset, we employ a housing price data from AnJuKe\footnote{https://shanghai.anjuke.com/}, an online platform for real estate sales and renting.  
Our data covers around 18K real estates at Shanghai in 2017\footnote{This data is publicly available: http://yangy.org/data.html}. 
Together with the GPS locations, we calculate the average housing price for a particular user's home, work place, and other active areas. 
Overall, the housing price in Shanghai spreads a wide range (Figure~\ref{fig:house}). 
For instance, in HuangPu area, the center of Shanghai, the average housing price has exceeded 100K yuan ($\sim$15K US dollars) per square meter. 
Meanwhile, other areas like MinHang has the housing price below 30K yuan per square meter.  
The average housing price in our dataset is 54.3K, with a standard deviation of 29.4K. 

\subsection{Computational Framework}
\label{sec:setup:framework} 
From a person's call logs, we can extract a mobile communication network, which can reasonably approximate a user's social network and how she develops her connections to others after moving to a new city. 
We can also obtain the geographical locations of users from our data, which is also valuable to understanding migrants' active areas in a new city.
We then formulate the following notations, which are consistent with our previous work~\cite{yang2018}. 

\para{Mobile communication networks.} 
Based on call logs, we establish a mobile communication network grouped by time periods.
Formally, we build a directed graph $G_t=(V_t, E_t)$ for a time period $t$, where $V_t$ is the set of users, and each directed edge $e_{ij} \in E$ indicates that $v_i$ calls $v_j$ during that time period ($v_i, v_j \in V_t$). 
Here $t$ can refer to a week or several days.
Note that only a subset of users in $V_t$ are labeled as \locals, \stays, or \leaves.

\para{Geographical locations.} 
For each call a person makes, we have access to the GPS location from the corresponding telecommunication tower.
We then group a person's locations by time periods.
We collect all the locations that a person makes calls in a time period $t$, and refer to this ordered list of locations for user $v$ as $L_v^t=[l_1, \ldots, l_n]$, where $l_i$ contains the latitude and the longitude.
We have geographical locations for the subset of users with labels since they are all users of China Telecom by definition.


\begin{figure*}[t]
\centering
\subfigure[Fraction of simiarly aged contacts. \label{fig:social_age}] {
\epsfig{file=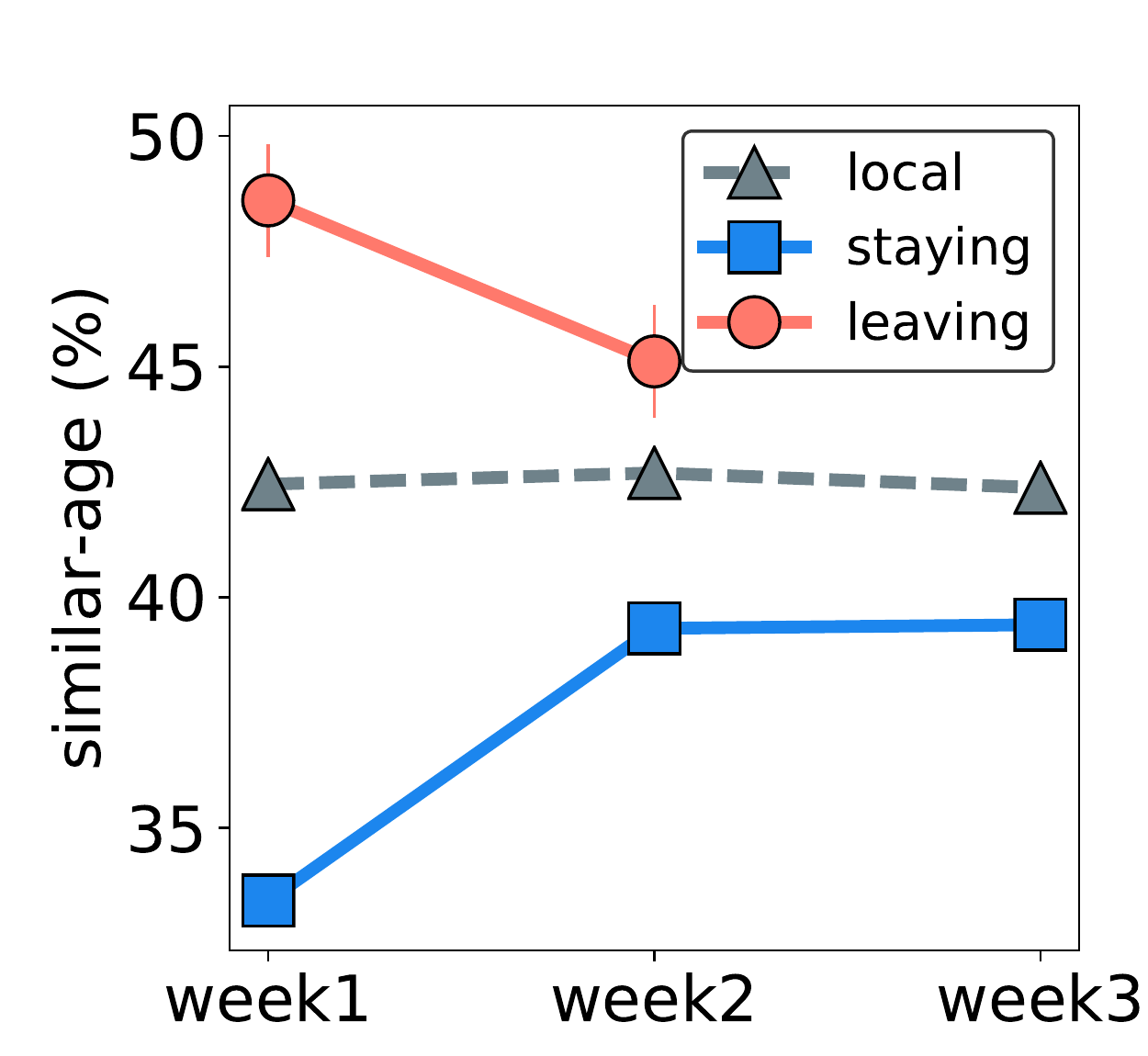, width=0.23\textwidth}
}
\subfigure[Fraction of same sex contacts.\label{fig:social_sex}] {
\epsfig{file=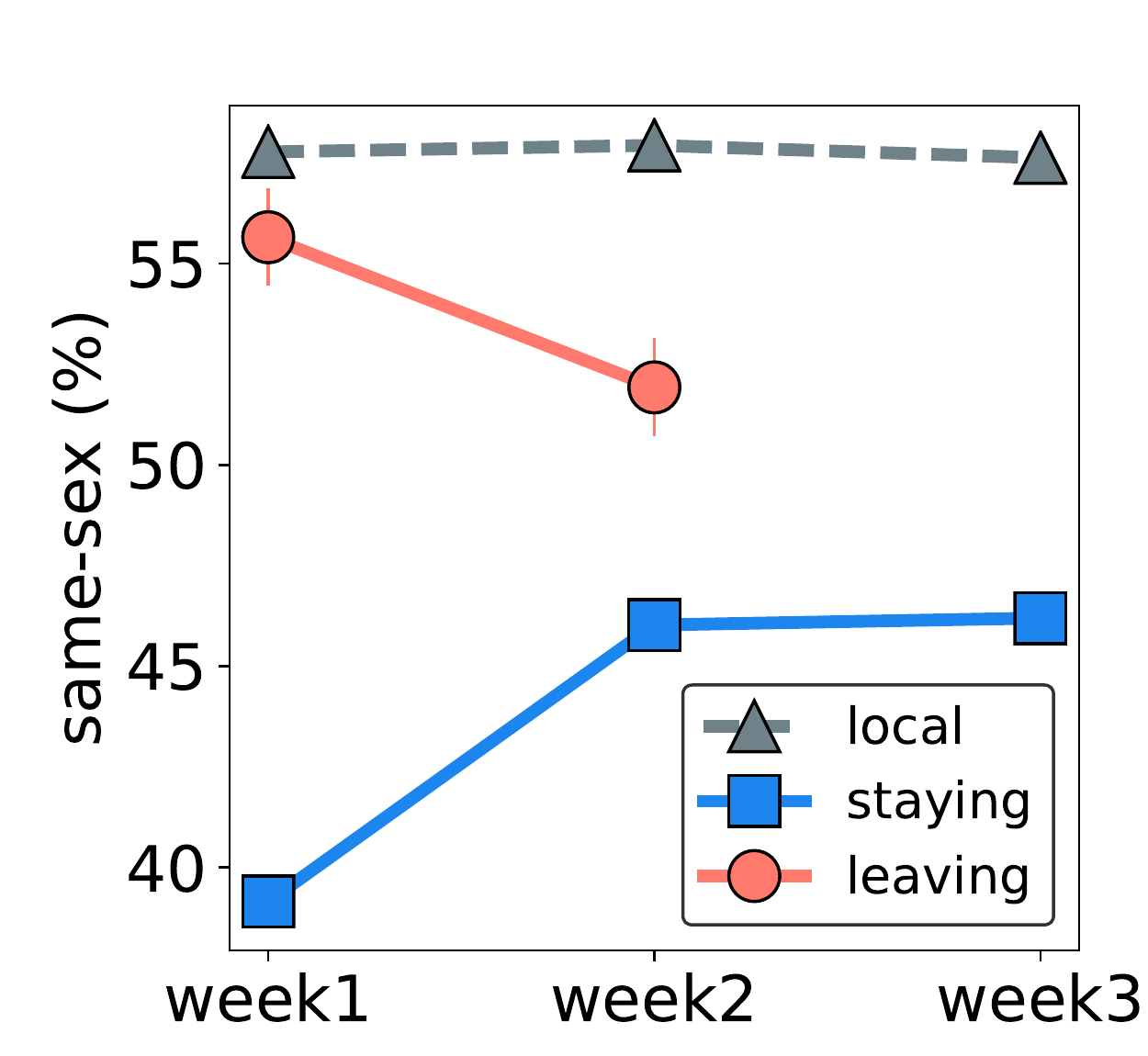, width=0.23\textwidth}
}
\subfigure[Degree.\label{fig:social_degree}] {
\epsfig{file=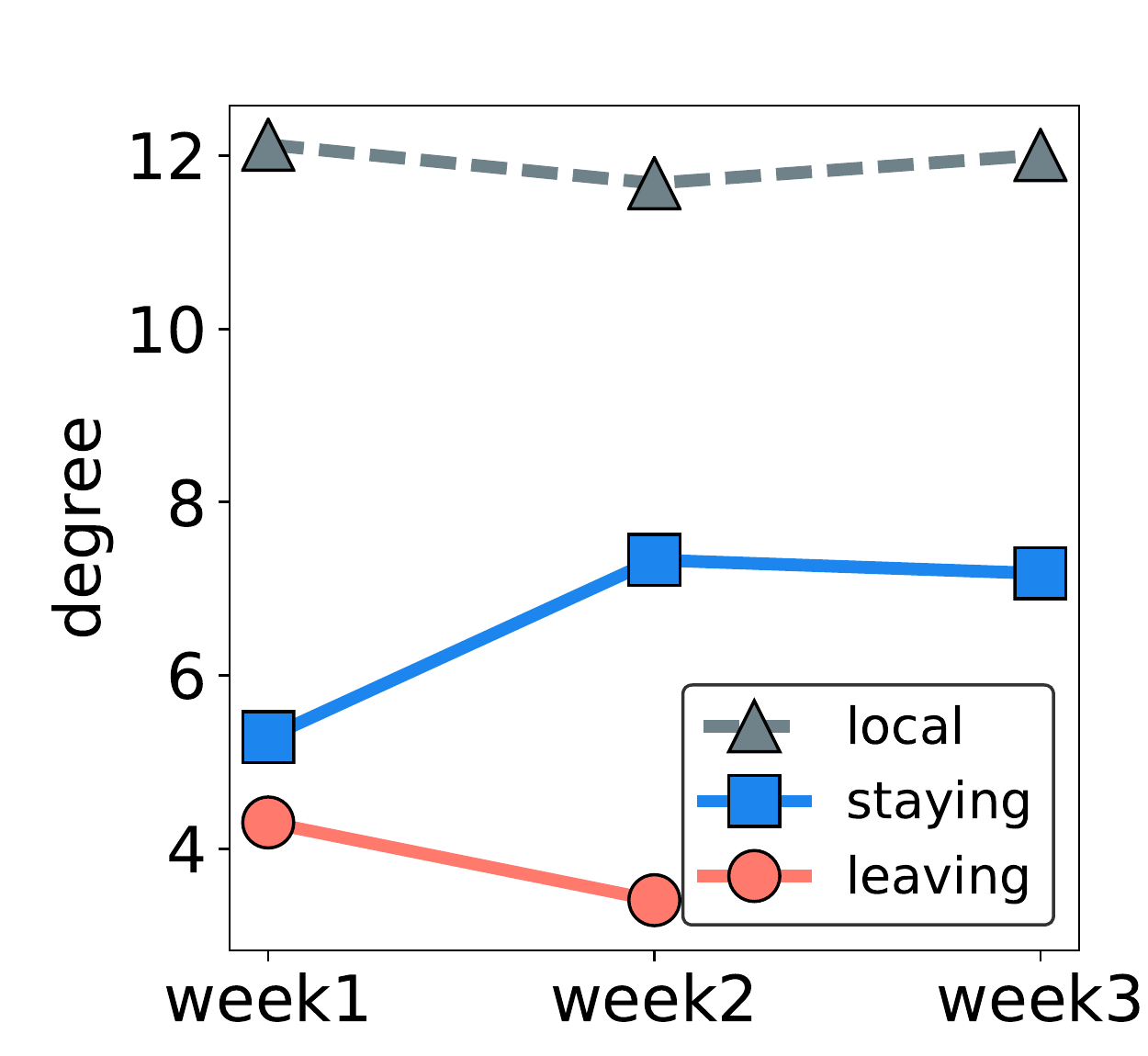, width=0.23\textwidth}
}
\subfigure[Average degree of contacts.\label{fig:social_contact_degree}] {
\epsfig{file=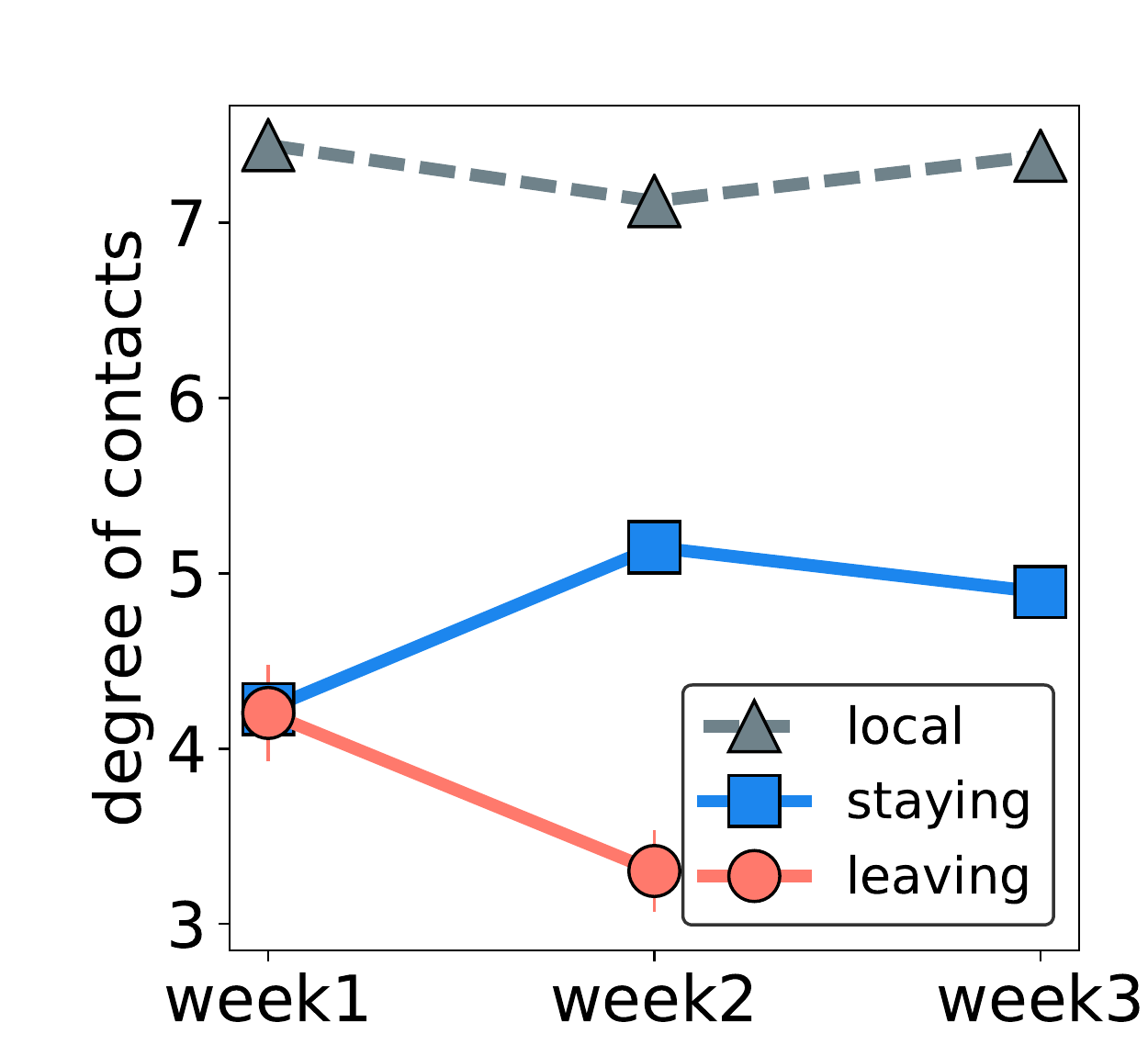, width=0.23\textwidth}
}
\\
\subfigure[Fraction of townspeople.\label{fig:social_townsman}] {
\epsfig{file=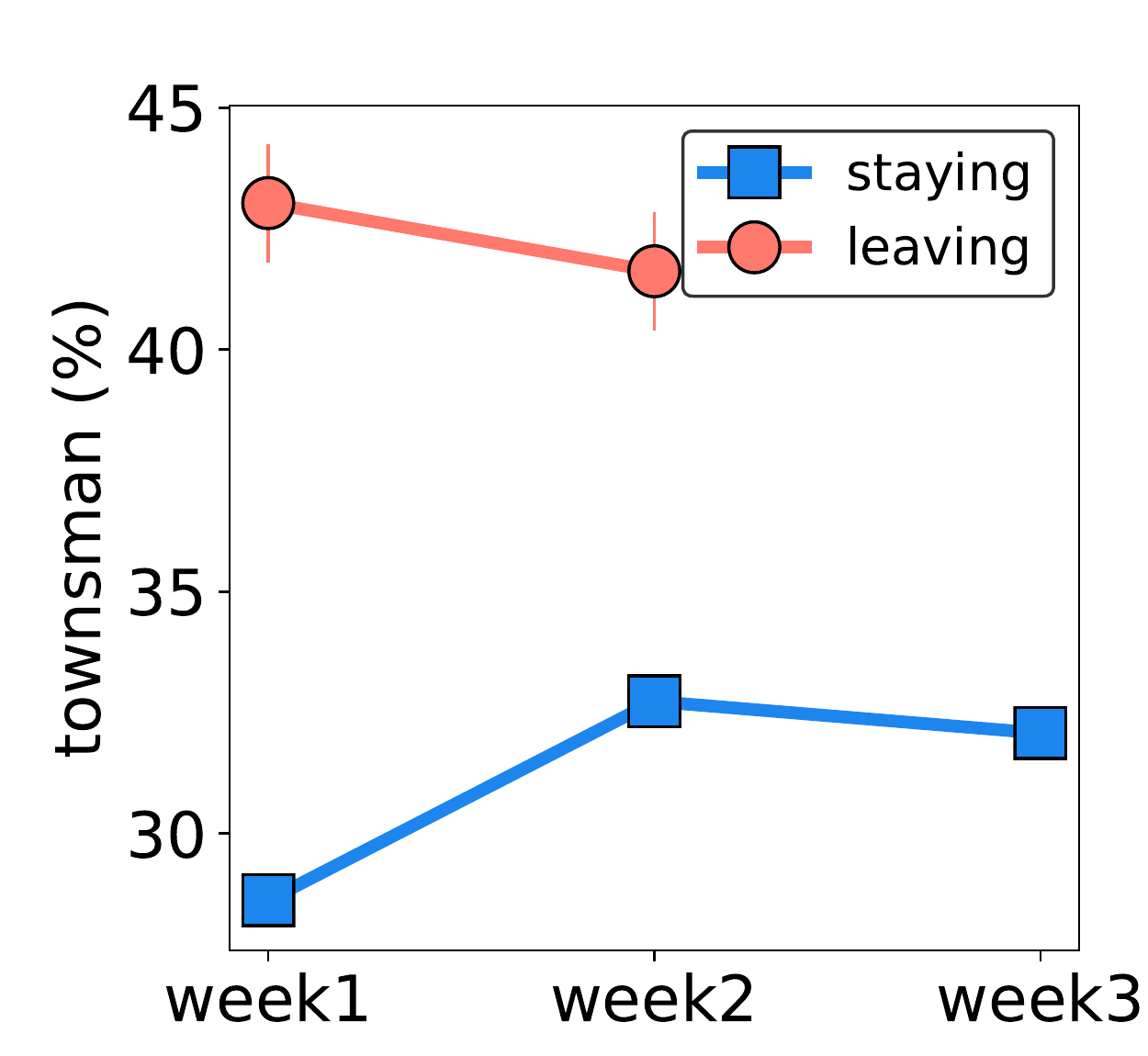, width=0.23\textwidth}
}
\subfigure[Province diversity.\label{fig:social_provdiv3}] {
\epsfig{file=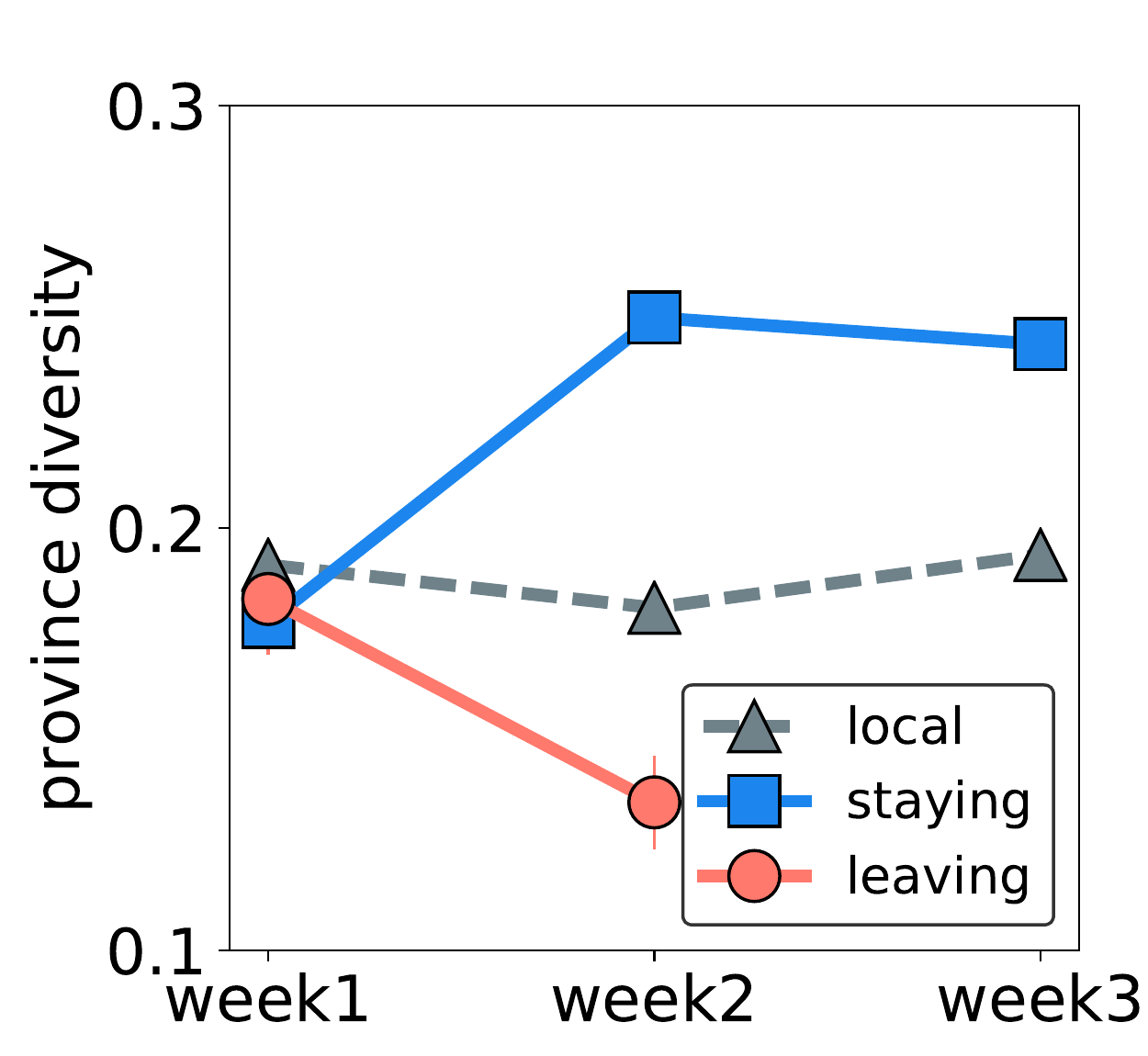, width=0.23\textwidth}
}
\subfigure[Clustering coefficient.\label{fig:social_cc}] {
\epsfig{file=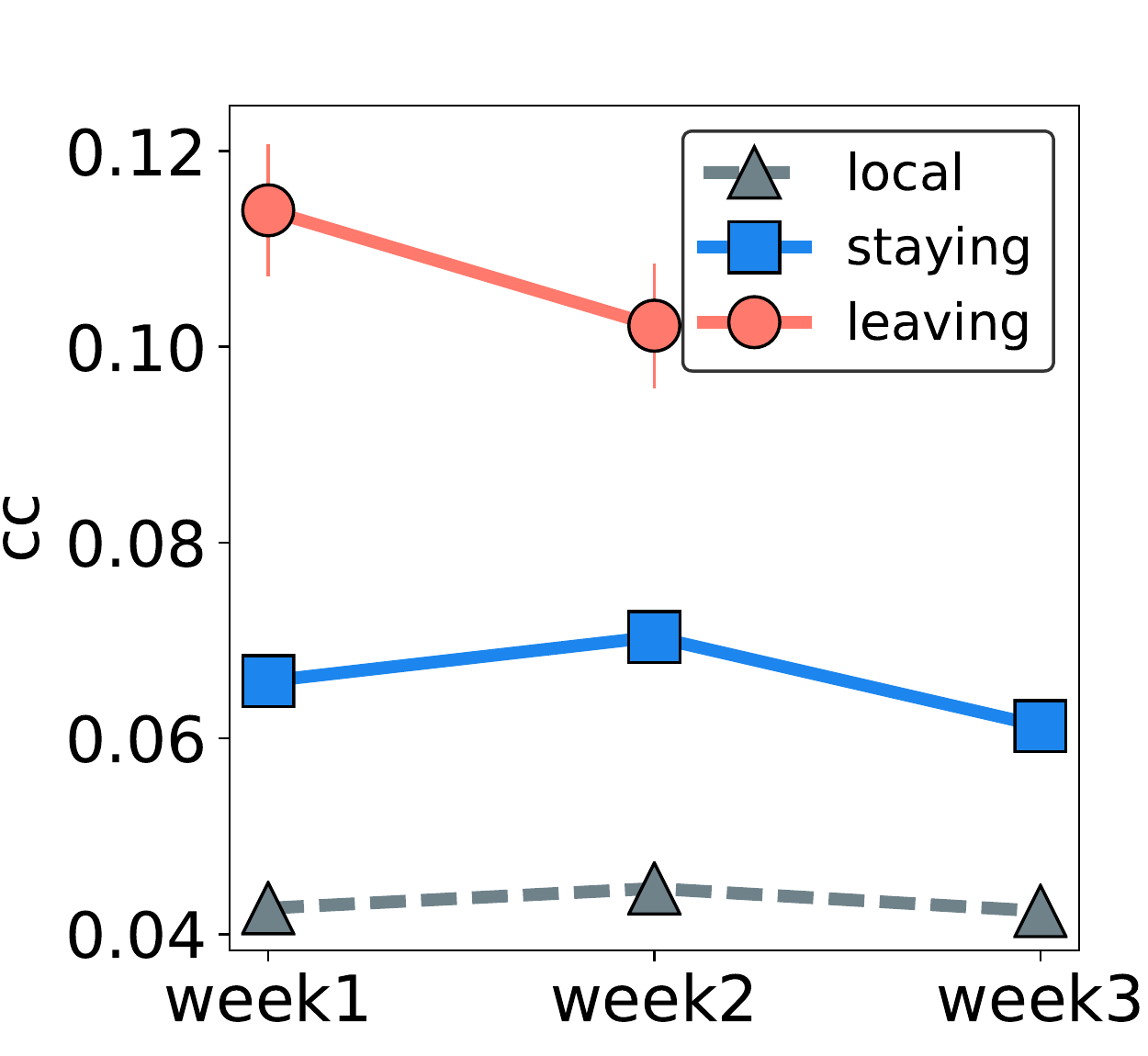, width=0.23\textwidth}
}
\subfigure[Communication diversity.\label{fig:social_diversity}] {
\epsfig{file=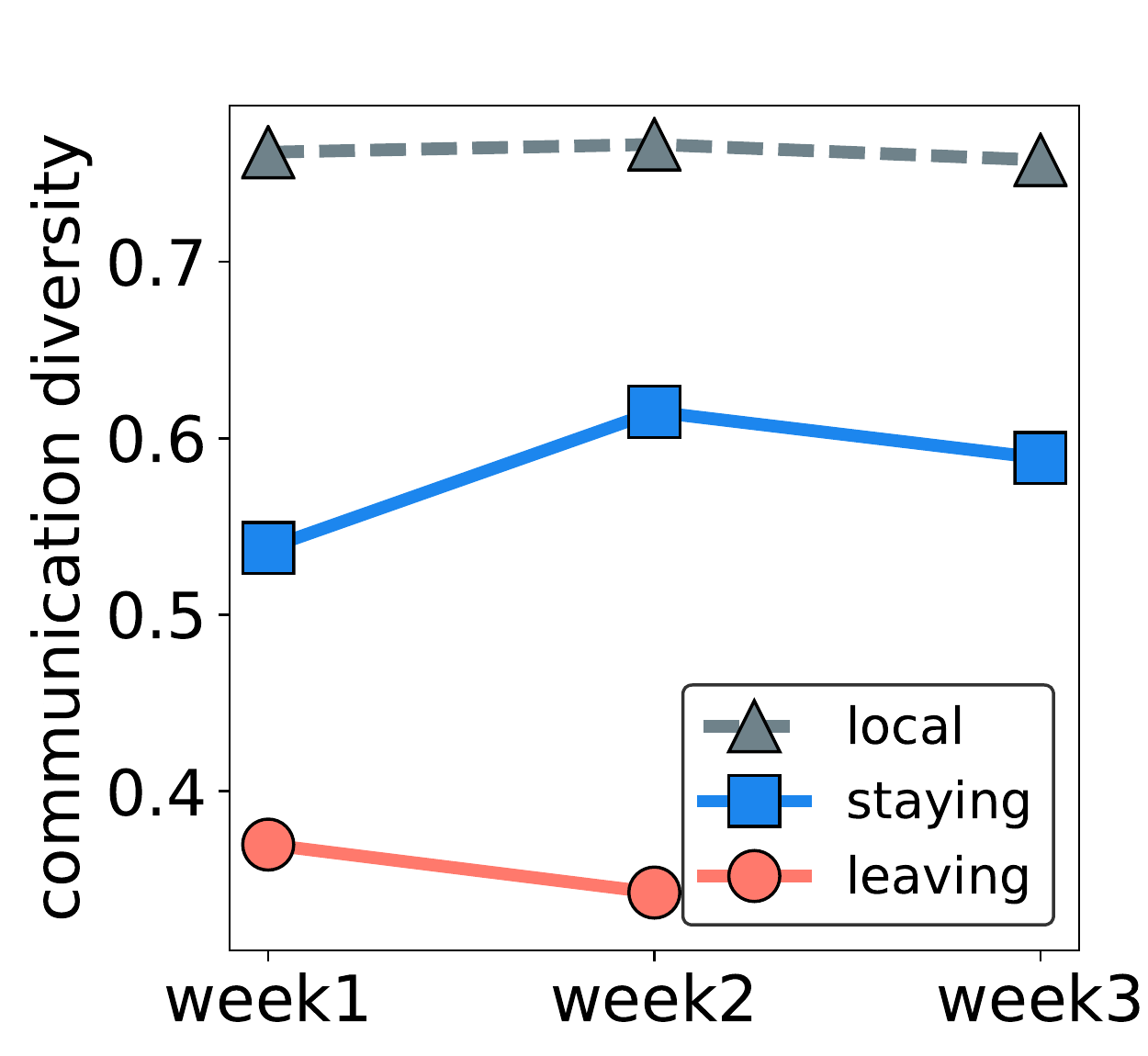, width=0.23\textwidth}
}
\caption{How \locals, \stays, and \leaves build social connections in the first three weeks. y-axis represents feature values based of an individual's ego-network and x-axis represents time. 
} \label{fig:social_feature}
\end{figure*}

\begin{figure}[t]
\centering
\epsfig{file=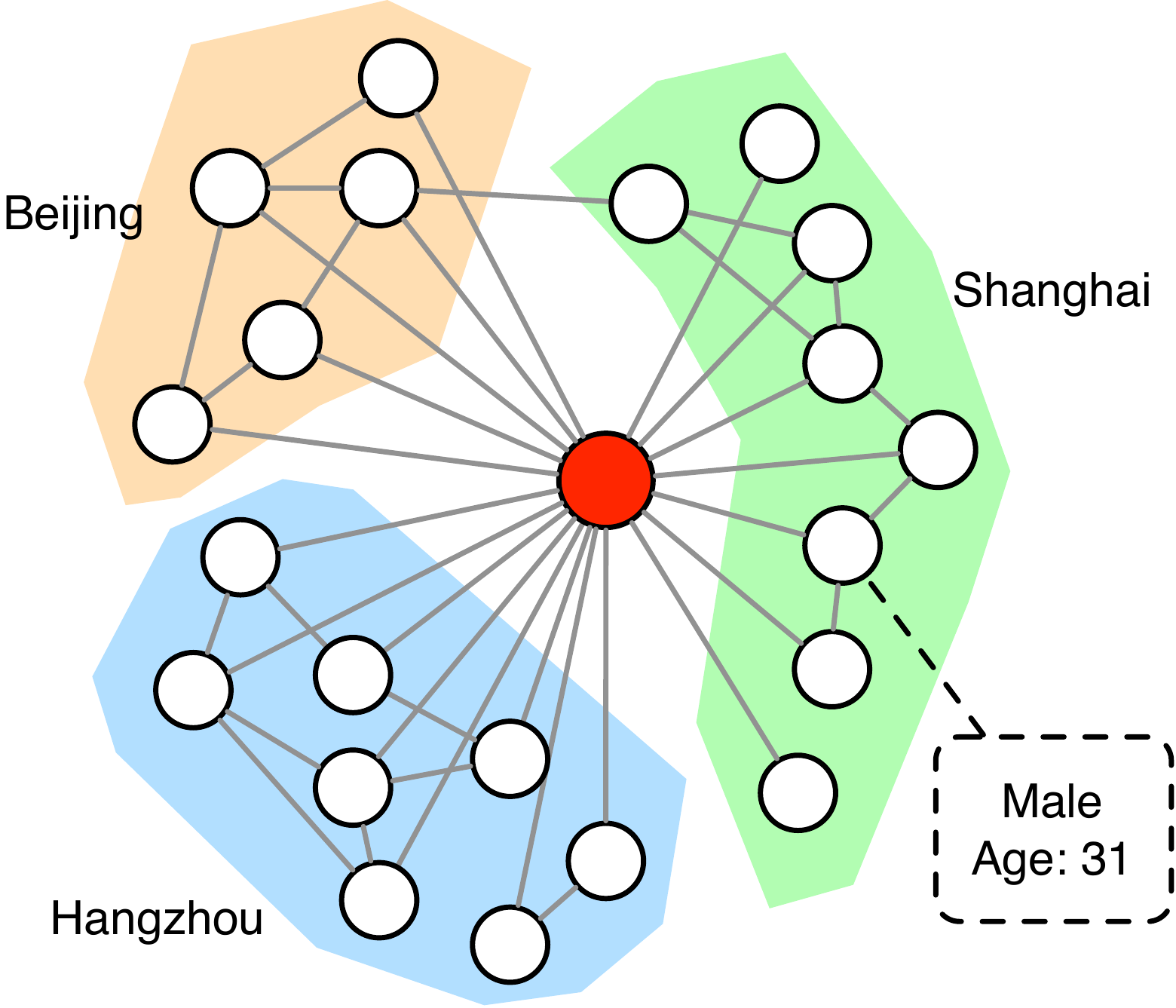, width=2in}
\caption{Example of an ego-network centered by a particular user $v$. 
\small We study how the structure of its ego-network evolve over time, and how the evolution pattern of a new migrant is correlated with her decision to settle down or leave the city. 
\normalsize
\label{fig:example}
}
\end{figure}

\section{The (Dis)integration of Migrants}
\label{sec:analysis}

In this section, we study the integration process of migrants in the first weeks after moving to a new city, and the disintegration process for some migrants who left early. 
To do that, we examine a wide range of factors from people's mobile communication networks and geographical locations. 
We propose four types of features:
ego network properties, call behaviors, geographical patterns, and housing price information.
In order to understand the integration process, we use \locals as a comparison point.
Therefore, we examine the differences between \locals, \stays, and \leaves in each week and how the features evolve over time.
Here we focus on explaining the motivation and evolving patterns of each feature by itself, and will examine their prediction performance in Section~\ref{sec:exp}.
Please refer to Table~\ref{tb:feature} in the appendix for computational details of each feature.
Note that we do not have feature values for \leaves at week 3 because they left in the third week.

\begin{figure*}[t]
\centering
\subfigure[Difference between out-calls and in-calls.\label{fig:call_diff}] {
\epsfig{file=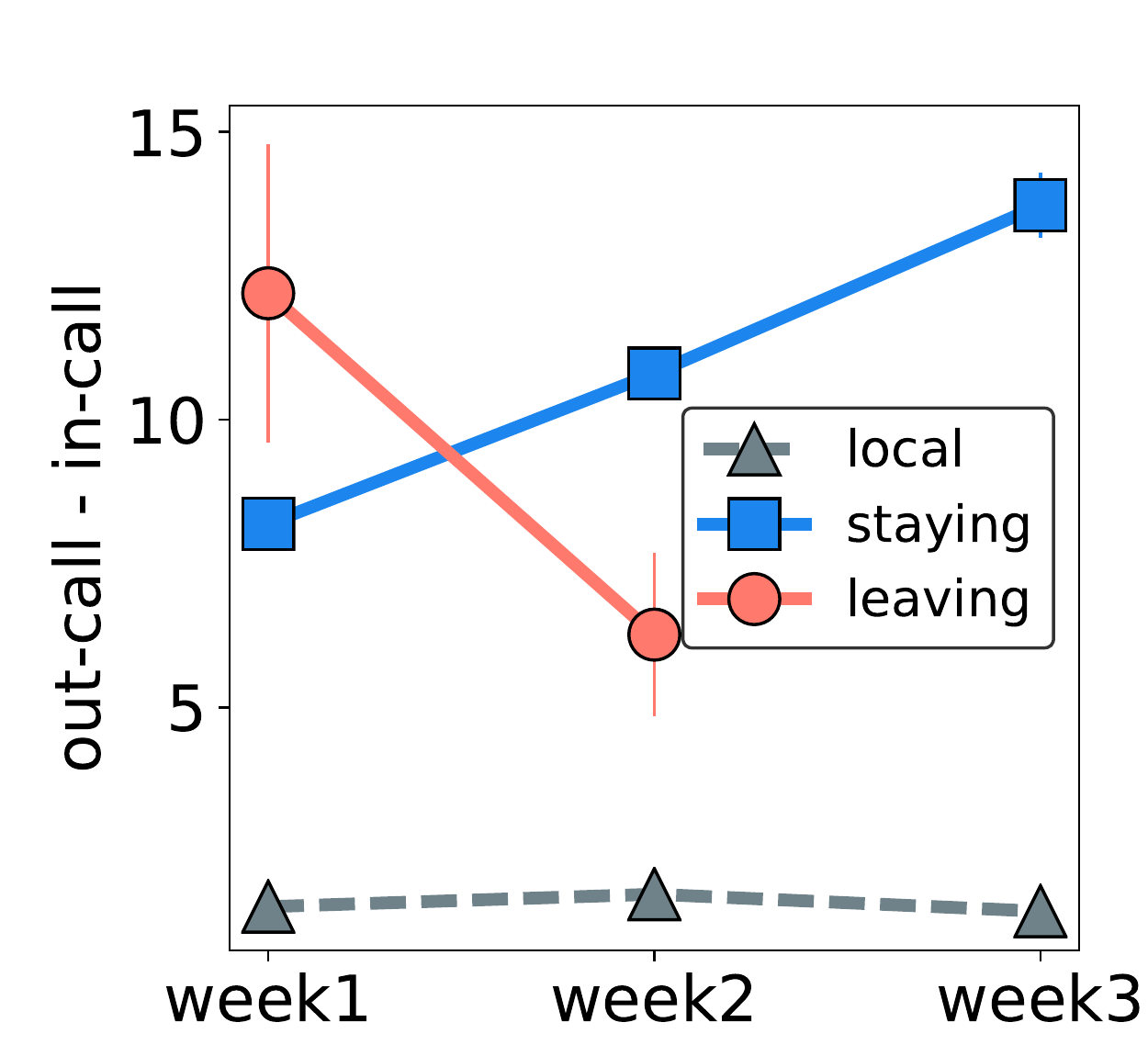, width=0.22\textwidth}
}
\subfigure[Average duration of calls. \label{fig:call_duration}] {
\epsfig{file=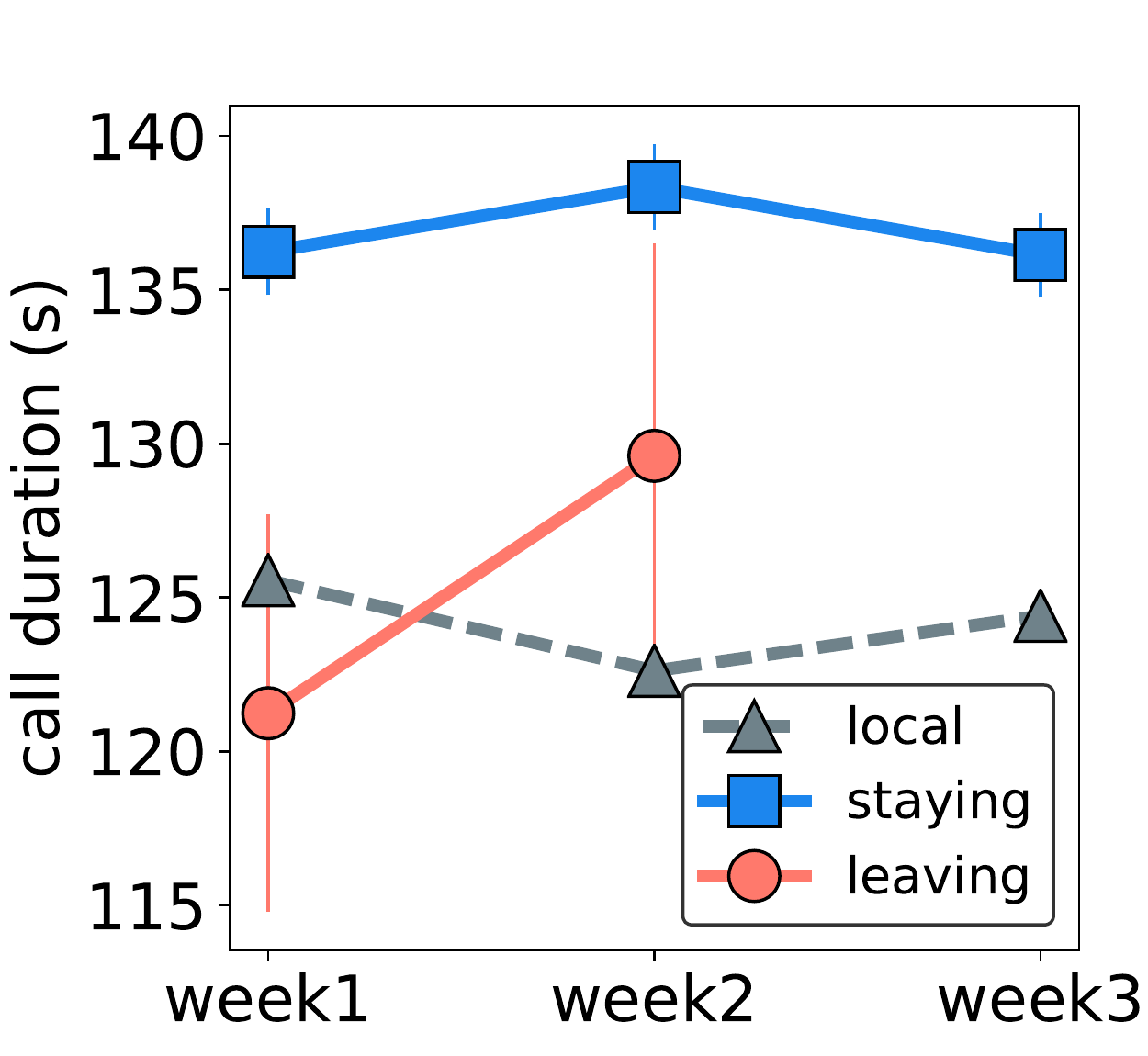, width=0.22\textwidth}
}
\subfigure[Average duration of calls made to locals. \label{fig:call_local_duration}] {
\epsfig{file=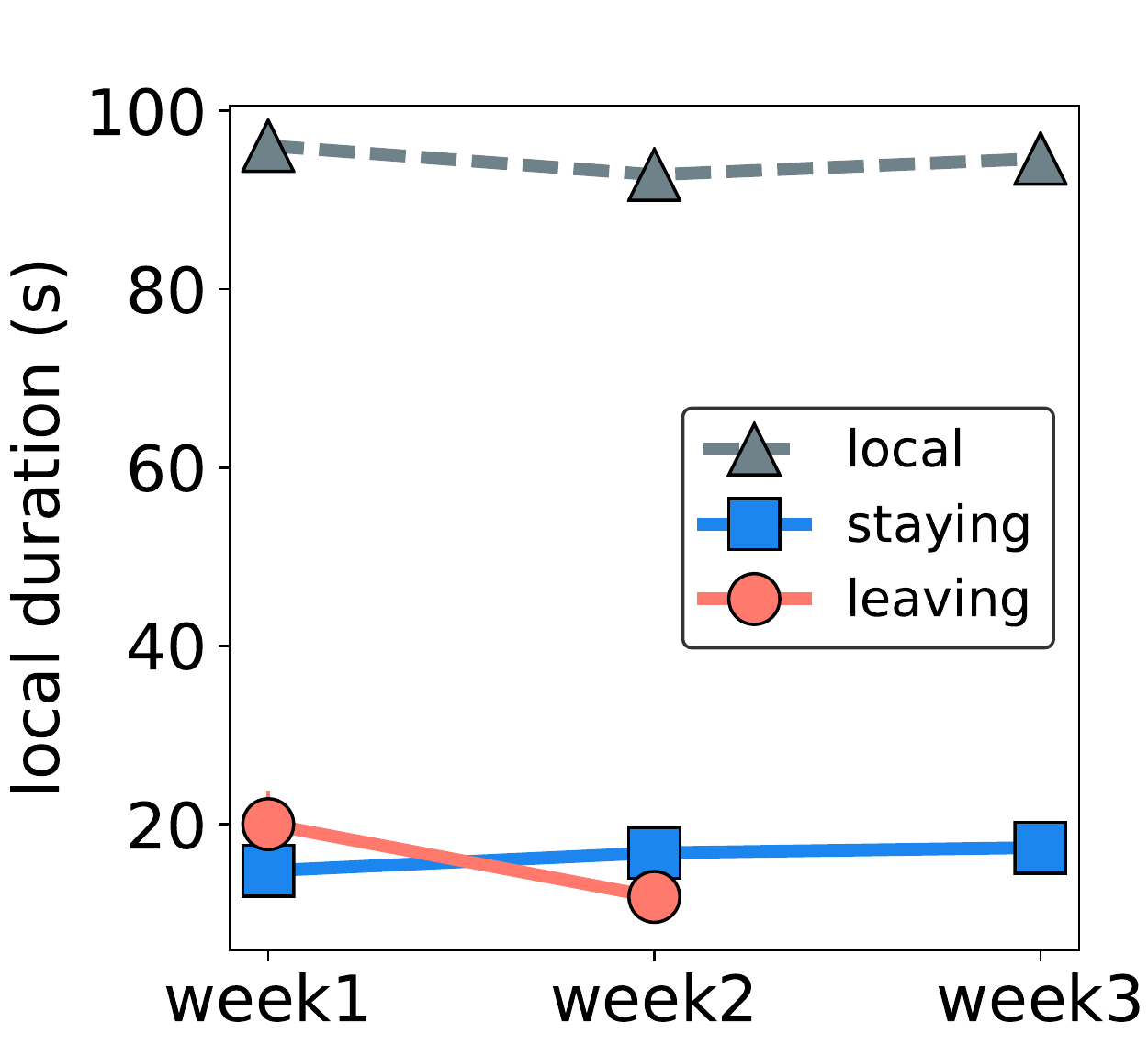, width=0.22\textwidth}
}
\subfigure[Fraction of reciprocal calls.\label{fig:call_reciprocal}] {
\epsfig{file=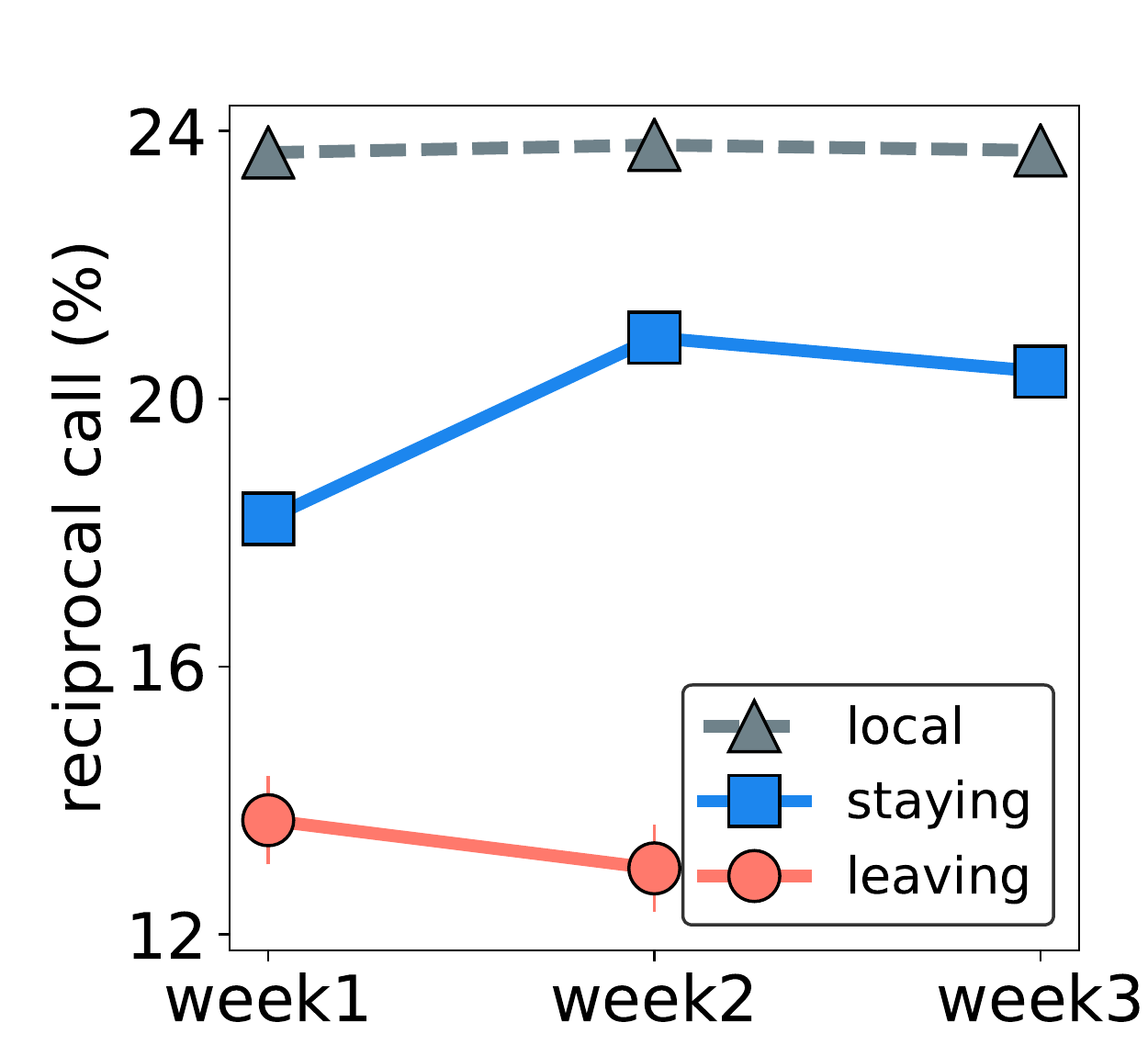, width=0.22\textwidth}
}
\caption{Call behavior of \locals, \stays, and \leaves.} \label{fig:call_feature}
\end{figure*}

\subsection{Ego network properties.} 
We first study 
how individuals build new connections and maintaining existing relationships in the first weeks after moving to a new city.
We extract features based on a person's ego network, the subgraph consisting of a person and all her neighbors~\cite{Freeman82:Ego}. 
We treat the telecommunication network as an undirected graph when building ego networks, so each neighbor of a person $v$ either called $v$ or received $v$'s calls.
As Figure~\ref{fig:example} illustrates,
we study characteristics of $v$ and her friends such as demographics, birthplaces, and connections to other people.  
Figure~\ref{fig:social_feature} presents the results. 
Overall, we find that the ego network features of \locals are more stable than those of new migrants over time. 

\vpara{Demographics.} 
Social homophily suggests that people tend to develop connections to those who are similar to themselves~\cite{mcpherson2001birds}. 
As Figure~\ref{fig:social_age} shows, \locals have a stable fraction of similarly-aged friends (around $0.41$), while \leaves have a larger fraction and \stays have a smaller fraction.
We see the results of \leaves and \stays are both getting closer to \locals, which indicates the integration process of migrants.  
This process seems to happen rather quickly in terms of talking to similar-aged friends.
As for sex (Figure~\ref{fig:social_sex}), \locals show the strongest homophily in sex, i.e., \locals have the largest fraction of contacts with the same sex. 
In comparison, new migrants who have a strong homophily in sex during the first week tend to be \leaves, while new migrants with more friends of the different sex tend to stay in Shanghai.

\vpara{Degree.}  
A people's degree reflects the number of contacts she has (Figure~\ref{fig:social_degree}(d)). 
As expected, \locals and their contacts have the largest degree.
In the first week, \stays and \leaves, and even their friends have very similar numbers of contacts.
However, \stays develop significantly more connections than \leaves in the second week. 
Such better connectedness also applies to the contacts that \stays make in the second week.

\vpara{Diversity of connections.} 
We finally examine the diversity of one's connections from three aspects: the birthplace of contacts, clustering coefficient, and communication diversity over contacts.

We analyze the birthplace of contacts both in the fraction of people who came from the same hometown and the diversity across different provinces.
\Leaves rely on people from the same hometown much more, while \stays start with a lower fraction of townspeople and grow a little bit over the integration (Figure~\ref{fig:social_townsman}). 
Remember that \locals were born in Shanghai, so they are not shown in this figure.
To further study this, we define a person $v$'s province diversity as the entropy of the distribution of birth provinces among $v$'s contacts, i.e., $-\sum_x p_x \log_2 p_x$, where $p_x$ is the probability that a contact of $v$ was born in province $x$. 
Figure~\ref{fig:social_provdiv3} again shows that \locals are pretty stable over time, while \stays have the most diverse group of contacts and \leaves have the lowest. 
This suggests that contacting people from different regions may help integrate in a new city. 

Clustering coefficient measures the fraction of triangles in the ego-network and indicates how likely a person's contacts know each other. 
From Figure~\ref{fig:social_cc}, we see that \leaves present the largest clustering coefficient, while \locals have the lowest. 
This suggests that new migrants that start with a close-knit group after moving to a big city may hinder themselves from integration.

Finally, inspired by the social diversity proposed in~\citet{eagle2010network}, we define the communication diversity as a function of Shannon entropy to quantify how a person split the number of calls to her friends, i.e., $\frac{-\sum_{j}p_{ij}\log(p_{ij})}{log(k_i)}$. 
Here $k_i$ is the out-degree and $p_{ij}$ is the probability defined as $p_{ij}=\frac{n_{ij}}{\sum_{l}n_{il}}$, where $n_{ij}$ is the number of calls user $v_i$ makes to user $v_j$. 
The result in Figure~\ref{fig:social_diversity}, again, suggests that developing more diverse connections may help new migrants integrate into a big city like Shanghai.


\begin{figure*}[t]
\centering
\subfigure[Moving distance.\label{fig:geo_move}] {
\epsfig{file=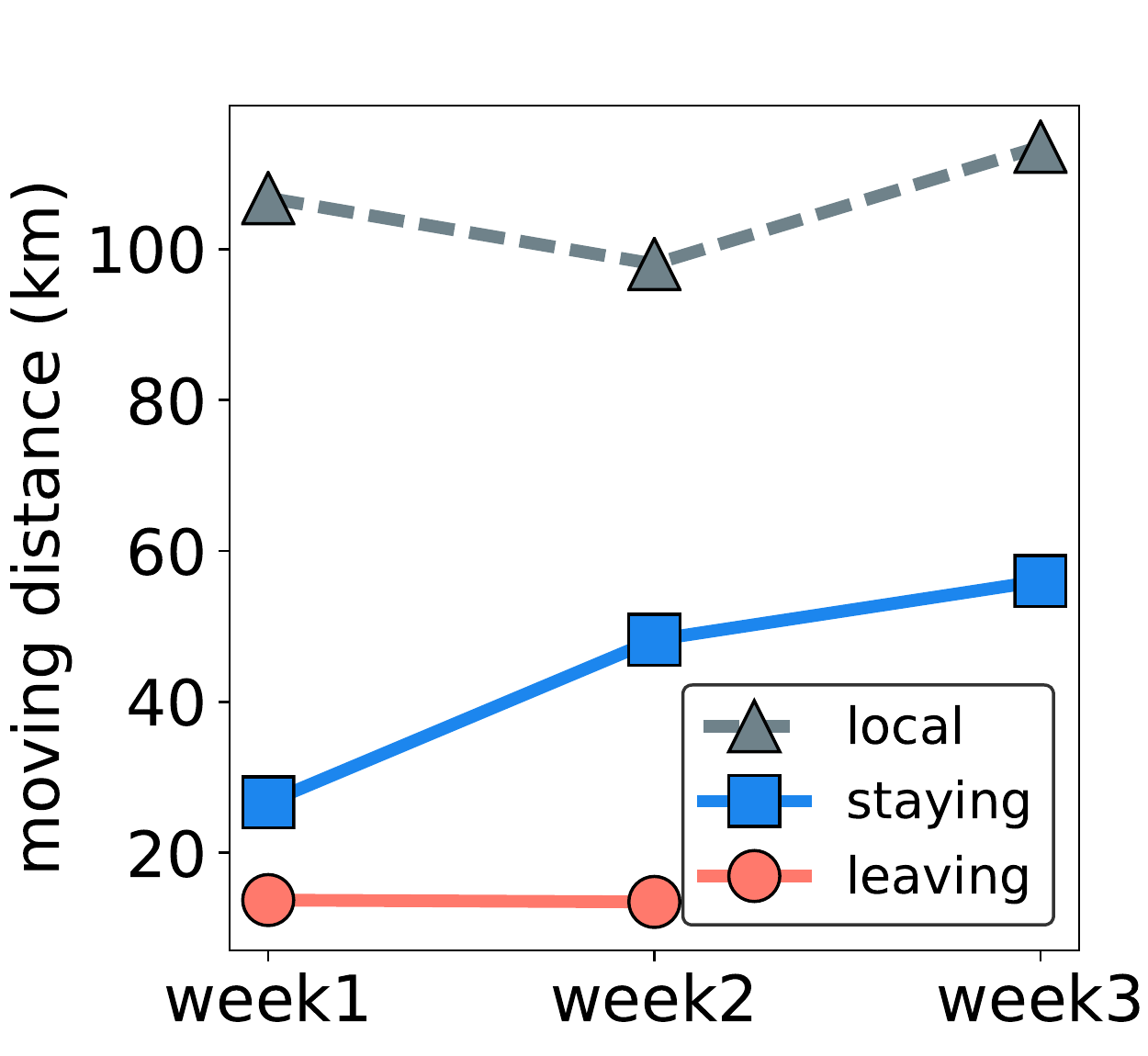, width=0.23\textwidth}
}
\subfigure[Average radius.\label{fig:geo_ave}] {
\epsfig{file=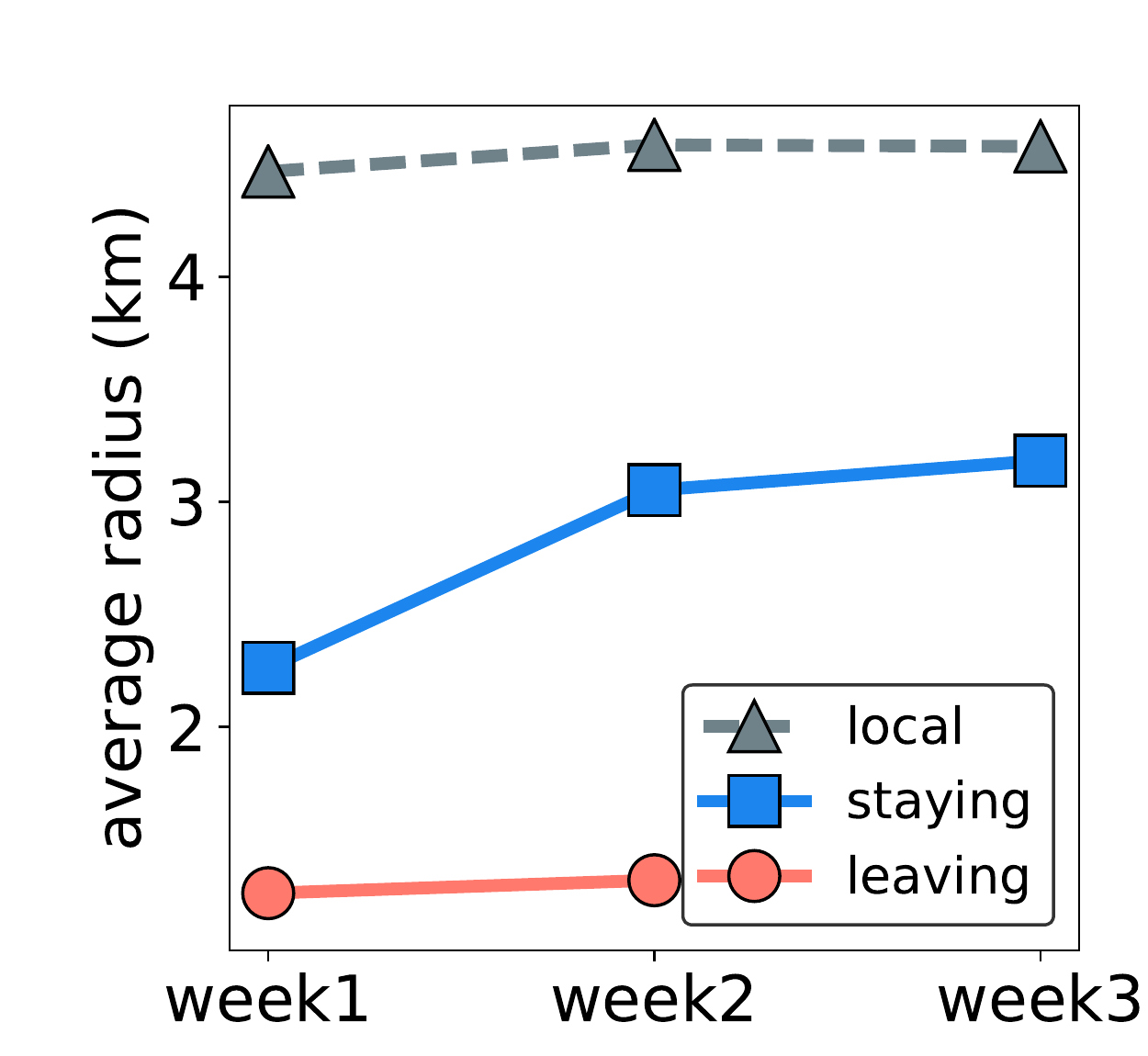, width=0.23\textwidth}
}
\subfigure[Max radius. \label{fig:geo_radius}] {
\epsfig{file=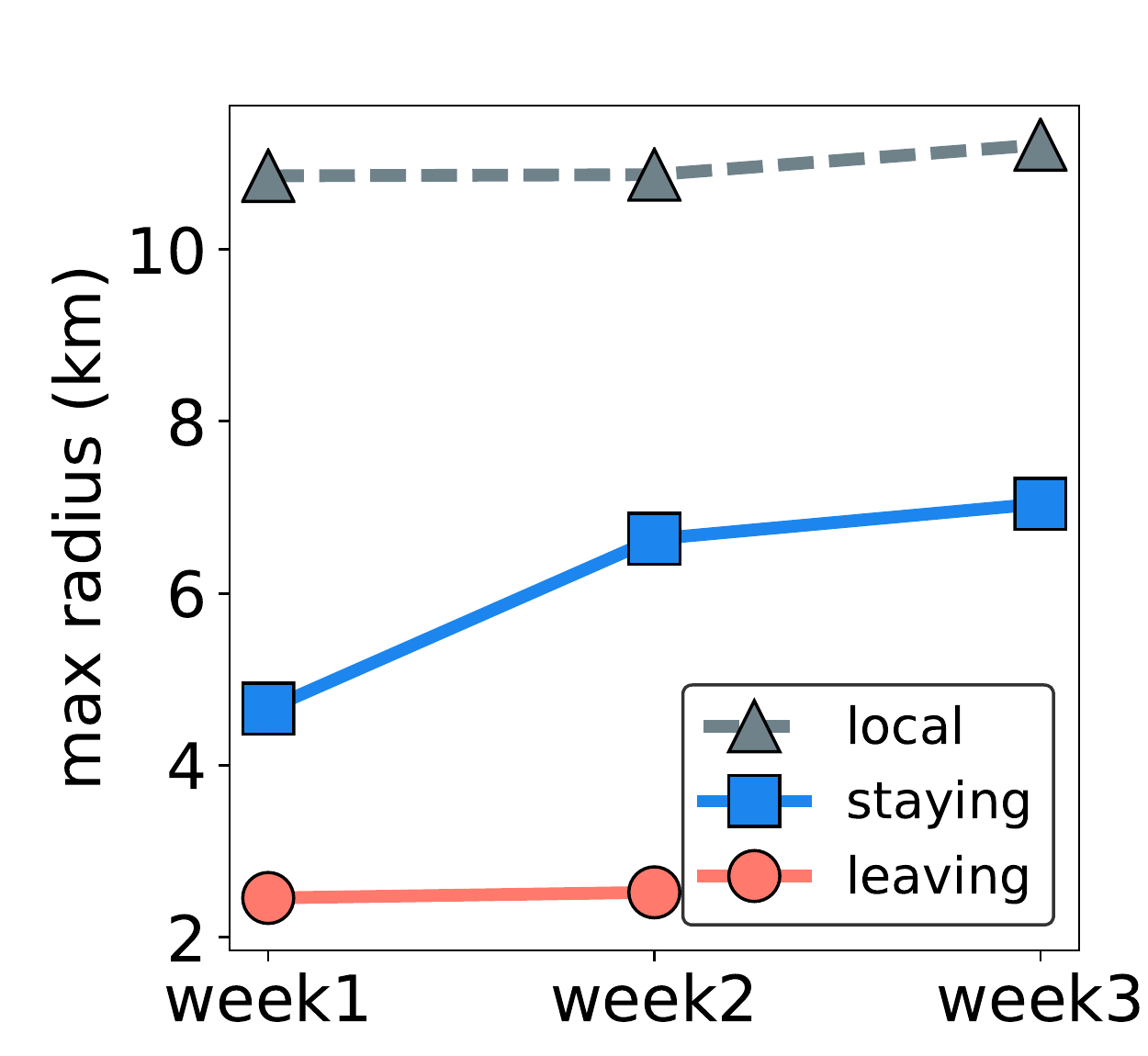, width=0.23\textwidth}
}
\subfigure[Distance between home and work palce. \label{fig:geo_distance}] {
\epsfig{file=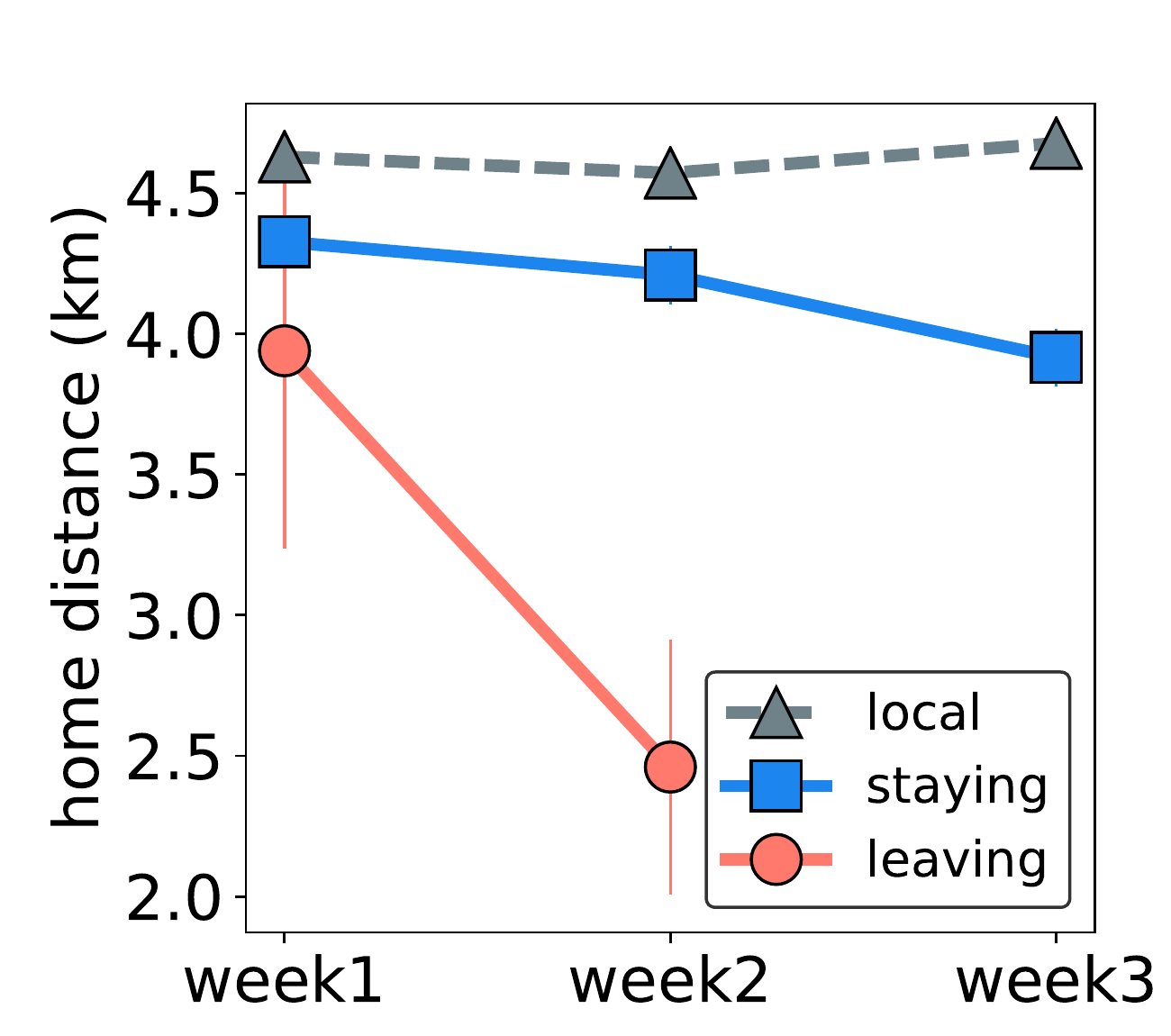, width=0.23\textwidth}
}
\caption{Geographical features of \locals, \stays, and \leaves (in kilometers).
} \label{fig:feature_geo}
\end{figure*}

\subsection{Call Behavior} 
For users' call behaviors, we first examine the difference between a person's outgoing calls and incoming calls in Figure~\ref{fig:call_diff}. 
The positive values suggest that 
the three groups of 
people in our dataset are more likely to make phone calls than to receive calls. 
We also see that the difference is larger for new migrants than \locals from the figure. 
Making more out-calls may be a sign that new migrants are developing initial connections.
\Stays make more out-calls as time grows, whereas \leaves make fewer.
Note that this feature of \stays grows more dissimilar to \locals, suggesting that two weeks is too short for \stays to integrate into \locals.  

The duration of calls may reflect the 
strength of a relation between two person.
Naturally, closed friends tend to have longer phone calls, while strangers are more likely to have a quick check-in. 
Figure~\ref{fig:call_duration} shows that \stays make much longer calls than \locals and \leaves.  
One possible explanation is that \stays need more time to figure out their initial life in a new city.
Lack of such relations, \leaves fail to integrate into Shanghai. 
Meanwhile, \locals, who have stable relations in Shanghai, are less likely to make long calls.
However, as Figure~\ref{fig:call_local_duration} shows, \locals make significant longer calls to other \locals, while new migrants do not develop strong relations with locals in the first three weeks after they arrive. 

Finally, we investigate the fraction of reciprocal calls (i.e., two-way relationships between users).
As Figure~\ref{fig:call_reciprocal} shows, \locals are more likely to have reciprocal relationships with their contacts, while the fraction of reciprocal calls is lower for \stays, and is lowest for \leaves.
This again shows that the personal networks of new migrants are still nascent. 
At week 2, \stays have a larger likelihood to have reciprocal relationships, whereas \leaves' likelihood decreases. 

\hide{
\begin{figure}[t]
\centering
\subfigure[\Leaves. \label{fig:leave_geo}] {
\epsfig{file=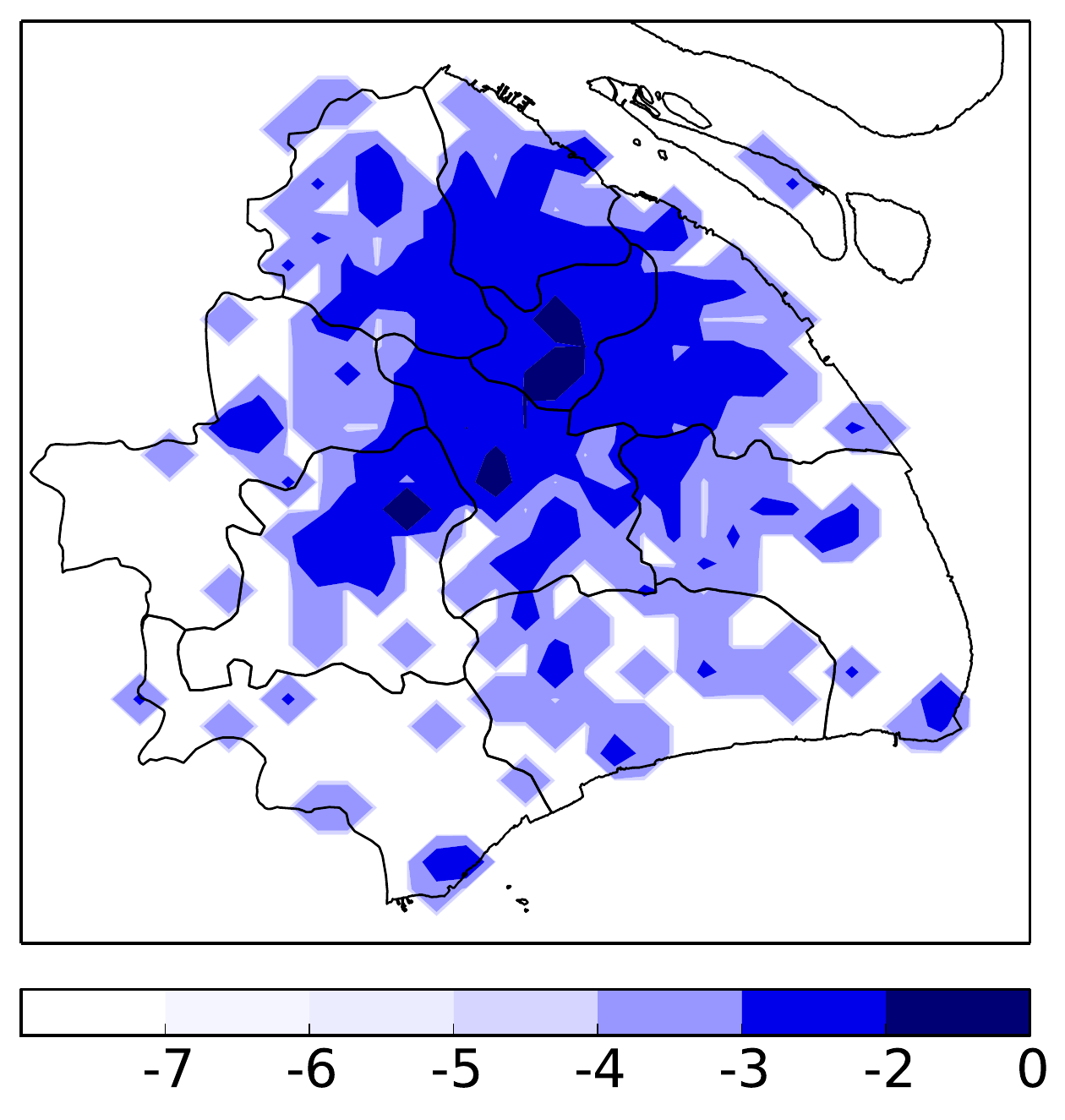, width=0.2\textwidth}
}
\hspace{-0.08in}
\hfill
\subfigure[\Stays.\label{fig:stay_geo}] {
\epsfig{file=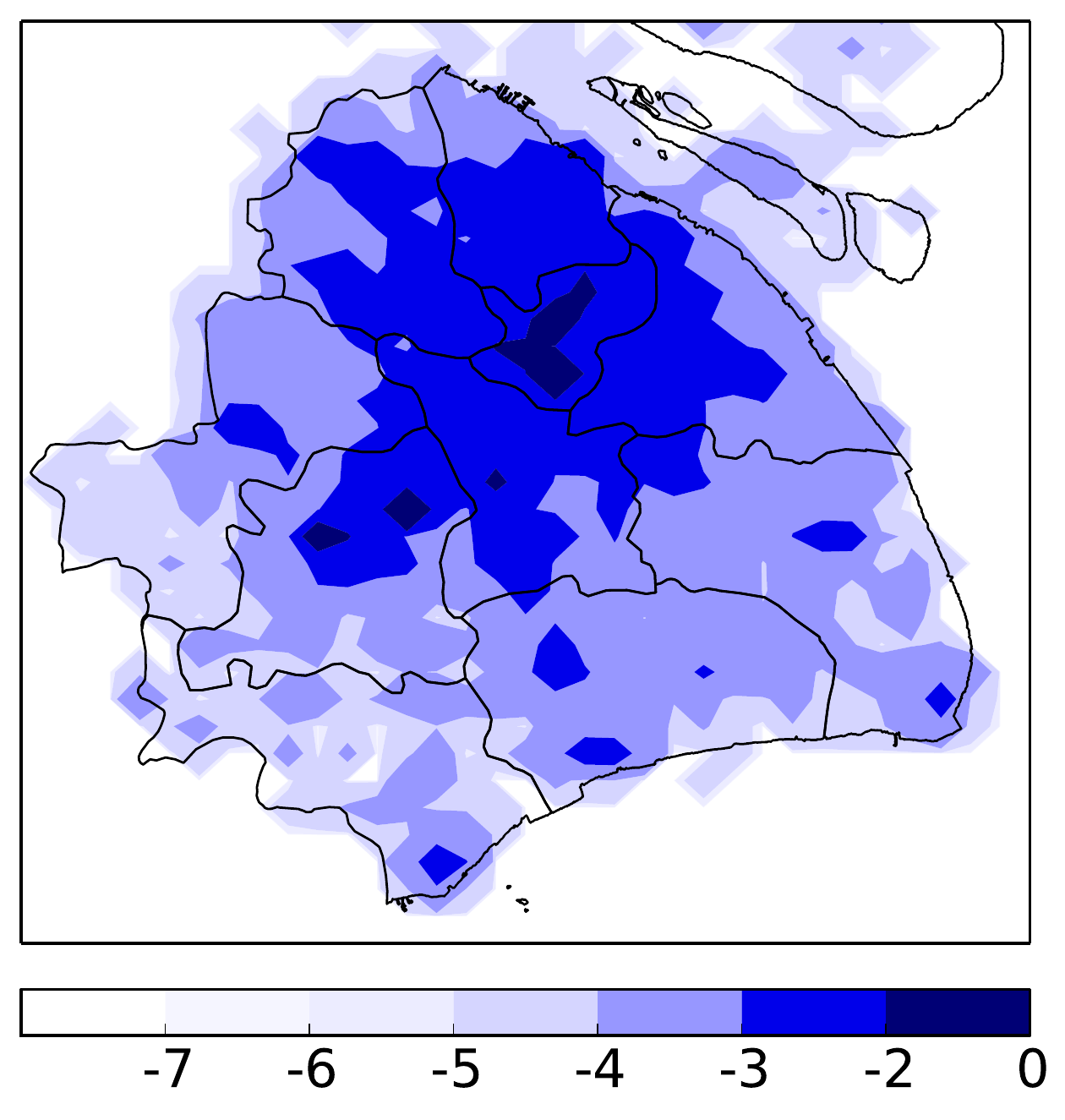, width=0.2\textwidth}
}
\caption{Geographical distributions of migrants. Each person is represented by the center of her active areas. The figures present the log probability of all users in each region.  
}
\label{fig:geographical}
\end{figure}
}

\subsection{Geographical Patterns}
We use locations to measure the mobility of migrants. 
Given a user $v$'s geographical locations $L_{v}^t=\{l_1, \cdots, l_n\}$, which is ordered by time and contains the latitude and the longitude of locations for user $v$ during a time period $t$, we can measure that user's active area from three different aspects.
First, we measure the {\em total distance} that the user moves as $\sum_i |l_i-l_{i-1}|$ (Figure~\ref{fig:geo_move}). 
Second, we compute a user $v$'s radius based on her center of mass $l_{CM}$ as $\frac{\sum_{l \in L_v^t}l}{|L_v^t|}$.
We define {\em average radius} as the average distance of $v$ from her center of mass, $\frac{\sum_{l\in L_v^t}|l-l_{CM}|}{|L_v^t|}$ (Figure~\ref{fig:geo_ave}). 
Similarly, we define {\em max radius} as the max distance of $v$ from her center of mass, i.e., $\max_{l \in L_v^t}|l-l_{CM}|$ (Figure~\ref{fig:geo_radius}). 
The results using these three statistics are consistent: \locals tend to move the most distance between calls on average and have much larger active area than new migrants. 
In comparison, \stays tend to expand their active areas, whereas \leaves' moving distance, average radius, and max radius only change slightly over time. 

By assuming that most people work in day time and go back home at night,
we can define a person $v$'s workplace as her center of mass during 9:00am to 16:00pm and her home as the center of mass during 20:00pm to 7:00am. 
We study the distances between a person's home and her workplace in Figure~\ref{fig:geo_distance}. 
\Locals have a stable distance over time and the small fluctuation can be explained by people's activity during the day and the night.
New migrants live slightly closer to workplace in the first week, while the distance becomes smaller over time, suggesting that new migrants may find new places to live after they obtain a job.
This decrease in distance between the day and the night can further alleviate the concern that these new migrants are temporary visitors.

\begin{figure}[t]
\centering
\subfigure[Average housing price of users' active areas. \label{fig:price_compare}] {
\epsfig{file=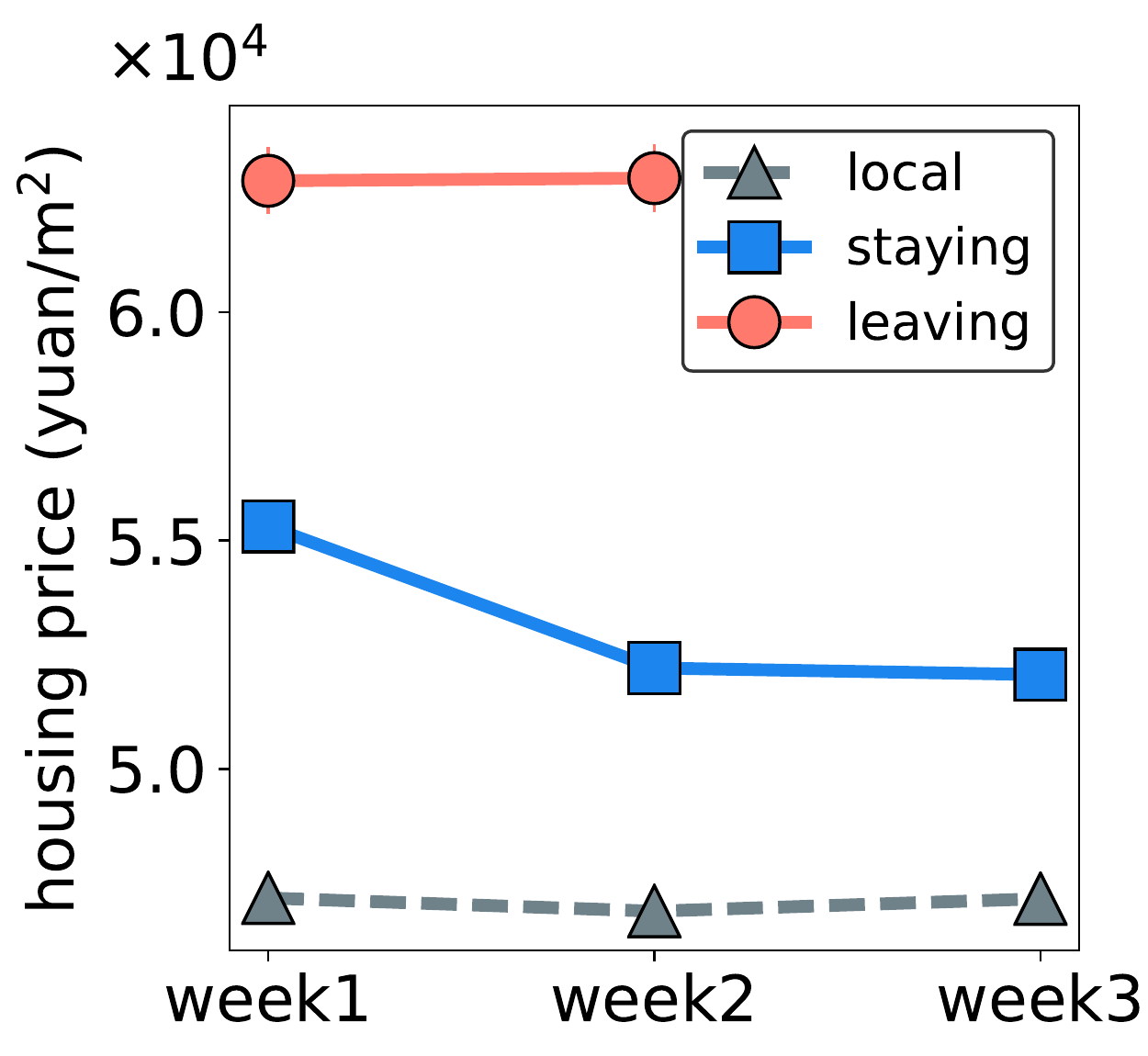, width=0.22\textwidth}
}
\hfill
\subfigure[Average housing price of friends' active areas.\label{fig:price_friend_hp}] {
\epsfig{file=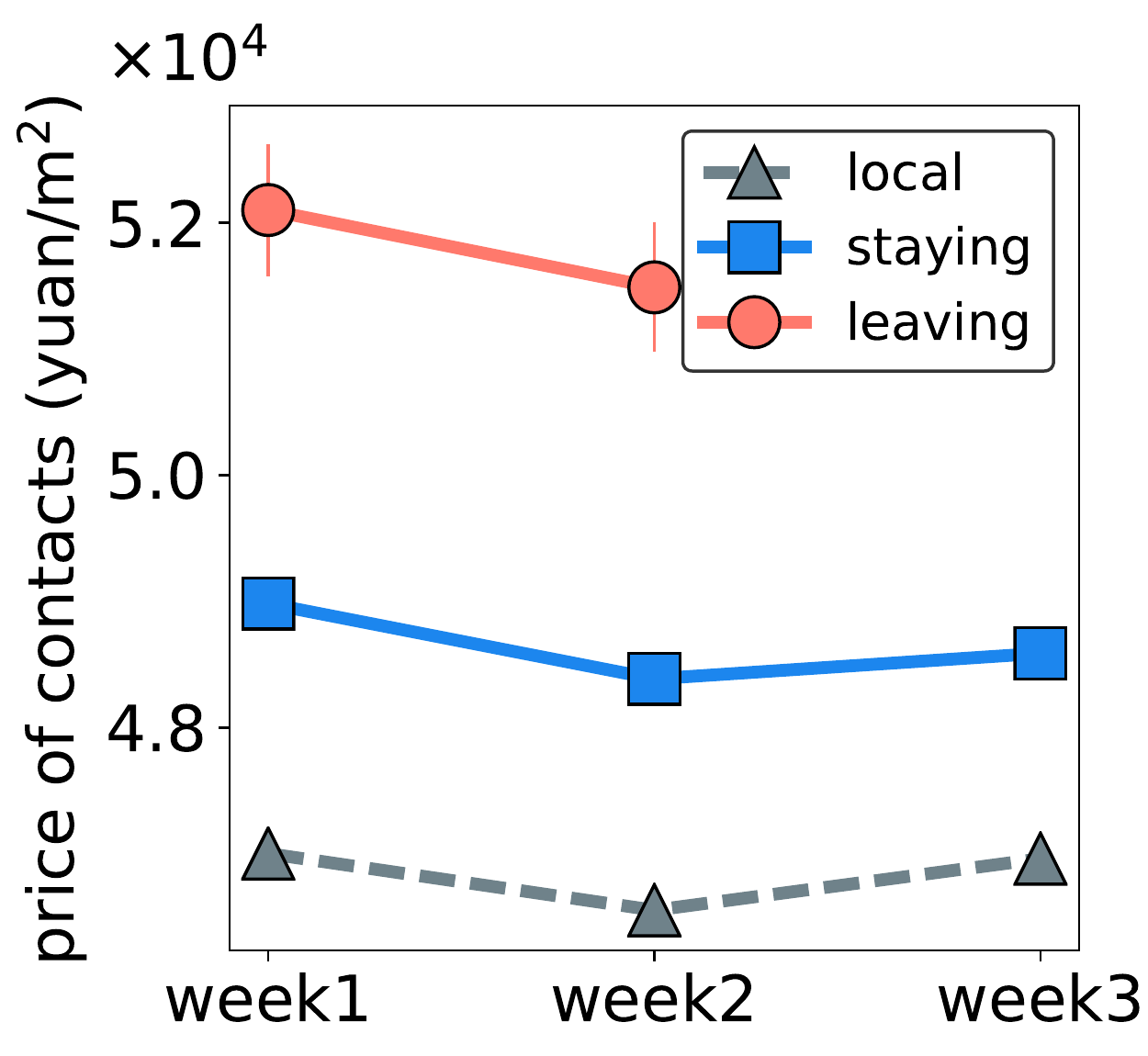, width=0.22\textwidth}
}
\caption{Housing price features of \locals, \stays, and \leaves.}
\label{fig:house_feature}
\end{figure}

\subsection{Housing Price Information}

The soaring housing price has been a central issue in the urbanization process of China \cite{Chen:HabitatInternational:2011,Wen:HabitatInternational:2015}.
Housing price can play an important role in a migrant's integration process.
Figure~\ref{fig:house} presents the overall distribution of housing price.
We compute the average housing price of a person's active geographical locations as well as of a person's friends' active locations. 

Surprisingly, we find that \locals tend to be active in the least expensive areas, while \leaves stay in the most expensive places (\figref{fig:price_compare}).
The average housing price of \stays drops significantly from the first week to the second week, but it is not the case for \leaves.
Similar results can be observed in the average housing price of a person's home. 
We omit the figure for space reasons. 
A new migrant may leave Shanghai early because she fails to find a place with a reasonable renting price. 
Due to social homophily, the average housing prices of one's active areas is similar to that of their friends (Figure~\ref{fig:price_friend_hp}). 

\vpara{Summary.} 
Through the evolving patterns of our proposed features, we find that 
\stays are able to move towards \locals in many dimensions, but three weeks is of course too short to finish the integration process.
However, 
comparing \stays to \leaves, \stays have more active and diverse mobile communication contacts and geographical movements. 
This suggests that actively expanding one's social network after moving to a new city is an important step for migrant integration.


\section{Predicting (Leaving) Migrants}
\label{sec:exp}

Having established the dynamic patterns of single features, 
we explore to what extent \locals, \stays, and \leaves are separable based on our proposed features. 
As these three groups of people have very different population sizes,
we set up two prediction tasks.
We first propose a binary classification task to predict if an individual is a \local or a new migrant,
and then work on distinguishing \leaves from \stays.
Both tasks are challenging due to the sparsity of data: less than 2\% people are new migrants and 4\% (1.5K) migrants left early in our dataset. 
The second task is more difficult as the behavior patterns of \leaves and \stays are more similar than those of new migrants and locals.  
However, accurate prediction of \leaves may allow for personalized service to help integration and insights from the second task can potentially inform urban policymakers, so we focus on the second task.
For both tasks, we use the same features listed in Table~\ref{tb:feature} in the appendix.

\begin{table}[t]
	\begin{tabular}{lrrr}
		\toprule
		Feature sets &  Precision & Recall & F1 \ \\ 
		\midrule
		all features  & 0.2355 & 0.8397 & {\bf 0.3678} \\ 
		ego network properties & 0.2097 & 0.8499 & 0.3363\\
		call behavior & 0.1021 & 0.8358 & 0.1820 \\ 
		geographical patterns & 0.0813 & 0.5971 & 0.1433\\ 
		housing price information &  0.0641 & 0.5347 & 0.1144 \\ 
		\midrule
		random guess & 0.0198 & 0.0198 & 0.0198 \\
		\bottomrule 
	\end{tabular}
	\caption{Distinguishing new migrants from locals using random forest with different set of features.}
	\label{tb:exp:migrant}
\end{table}

\subsection{Distinguishing New Migrants from Locals}
\label{sec:exp:migrant}
Our first binary classification task is to distinguish new migrants from locals. 
Formally, given a user $v$, a set of $v$'s mobile communication networks $\{G_t\}$ over the first 14 days since she moves to Shanghai ($1 \leq t \leq 14$), and the geographical locations $L_{v}^t$ of $v$ at time $t$ (the $t$-th day), our goal is to predict if $v$ is a new migrant or a local. 
We conduct 5-fold cross-validation and use precision, recall, and F1-score for evaluation, with the minority class, i.e., new migrants, as the target class. 


Table~\ref{tb:exp:migrant} demonstrate the results of random forest in this task.
Recall that in our dataset, there are 1.8 million locals and 35.5 thousand new migrants (52:1). 
Thus random guessing would obtain an F1-score around 0.02. 
Our method is able to outperform this random baseline significantly with an F1 of 0.36. 
We further compare the effectiveness of different feature sets by training a classifier that includes a single feature set and excludes the other feature sets. 
Table~\ref{tb:exp:migrant} shows that using every singe feature set outperforms random guessing.
Ego network properties performs the best, followed by geographical patterns. 

\begin{figure*}[t]
\centering
\subfigure[Precision. \label{fig:pred_precision}] {
\epsfig{file=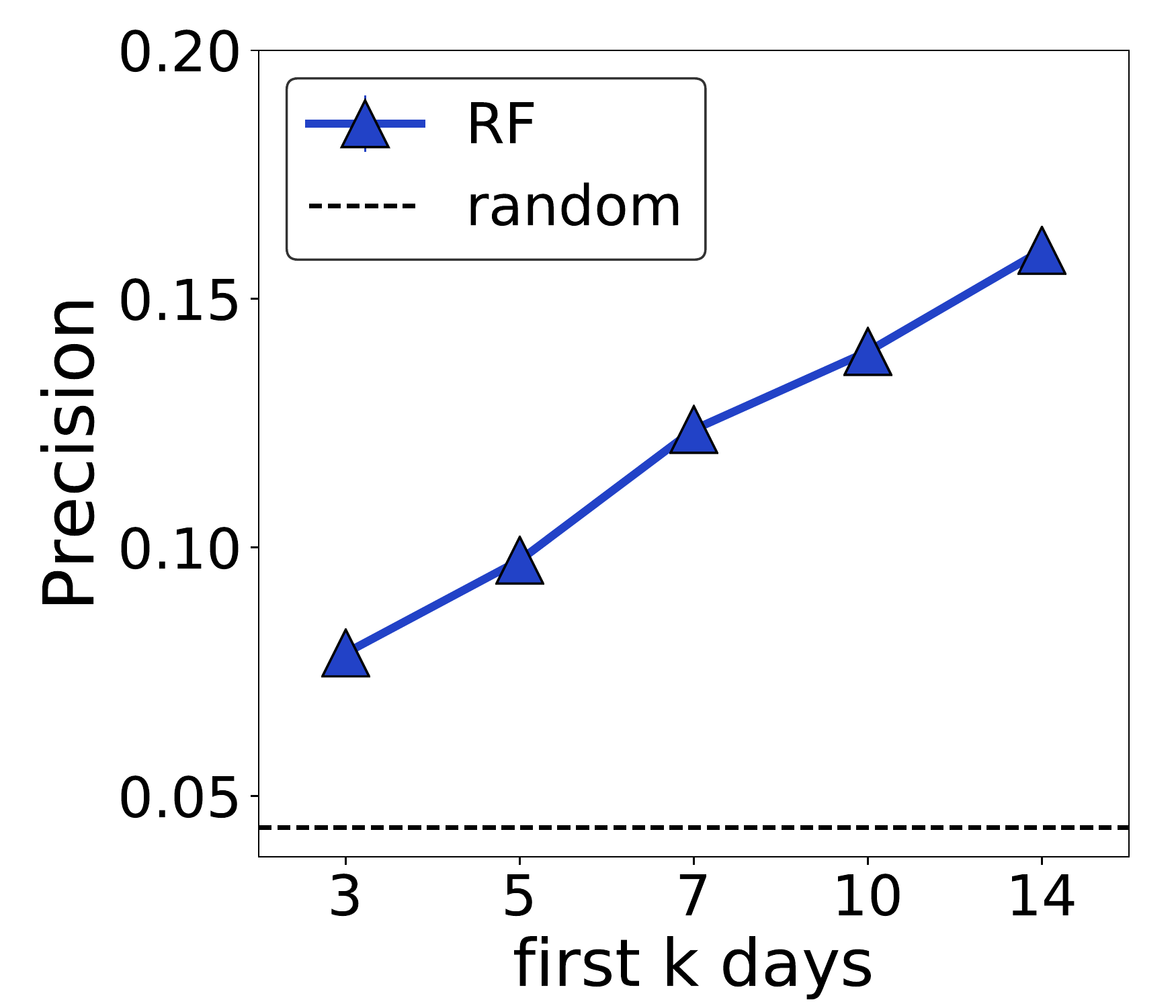, width=0.23\textwidth}
}
\hfill
\subfigure[Recall.\label{fig:pred_recall}] {
\epsfig{file=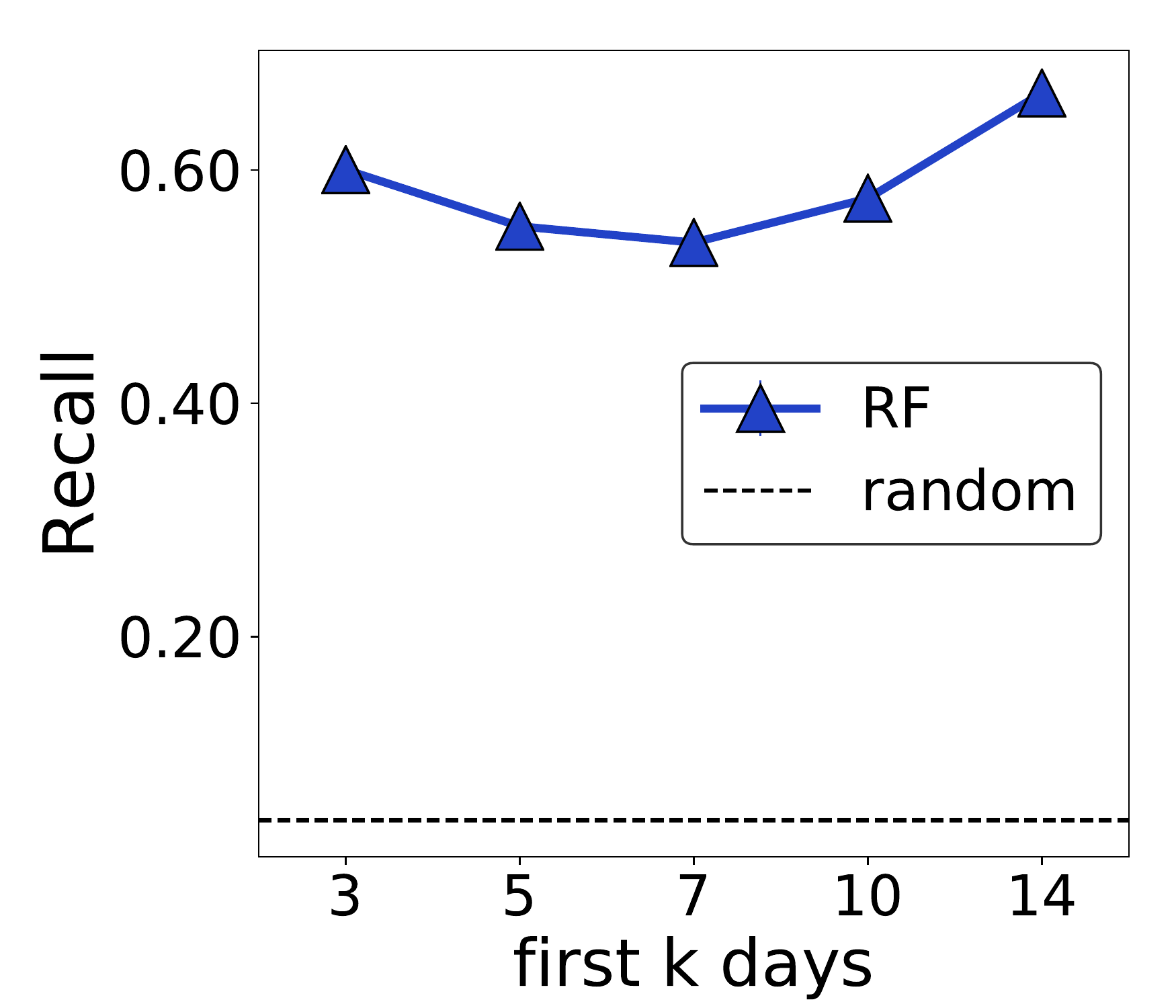, width=0.23\textwidth}
}
\hfill
\subfigure[F1.\label{fig:pred_f1}] {
\epsfig{file=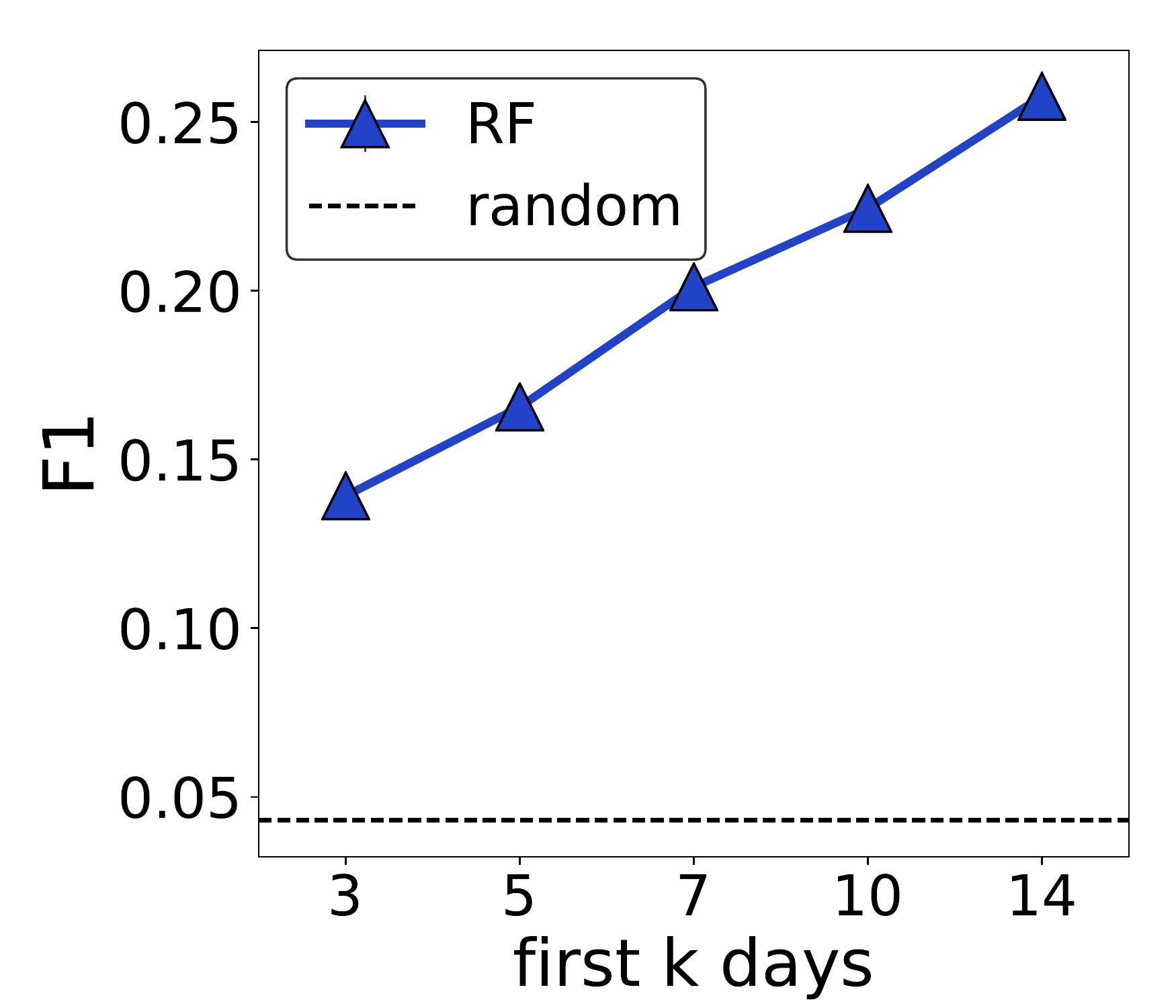, width=0.23\textwidth}
}
\hfill
\subfigure[Disentangling performance improvement.\label{fig:time}] {
\epsfig{file=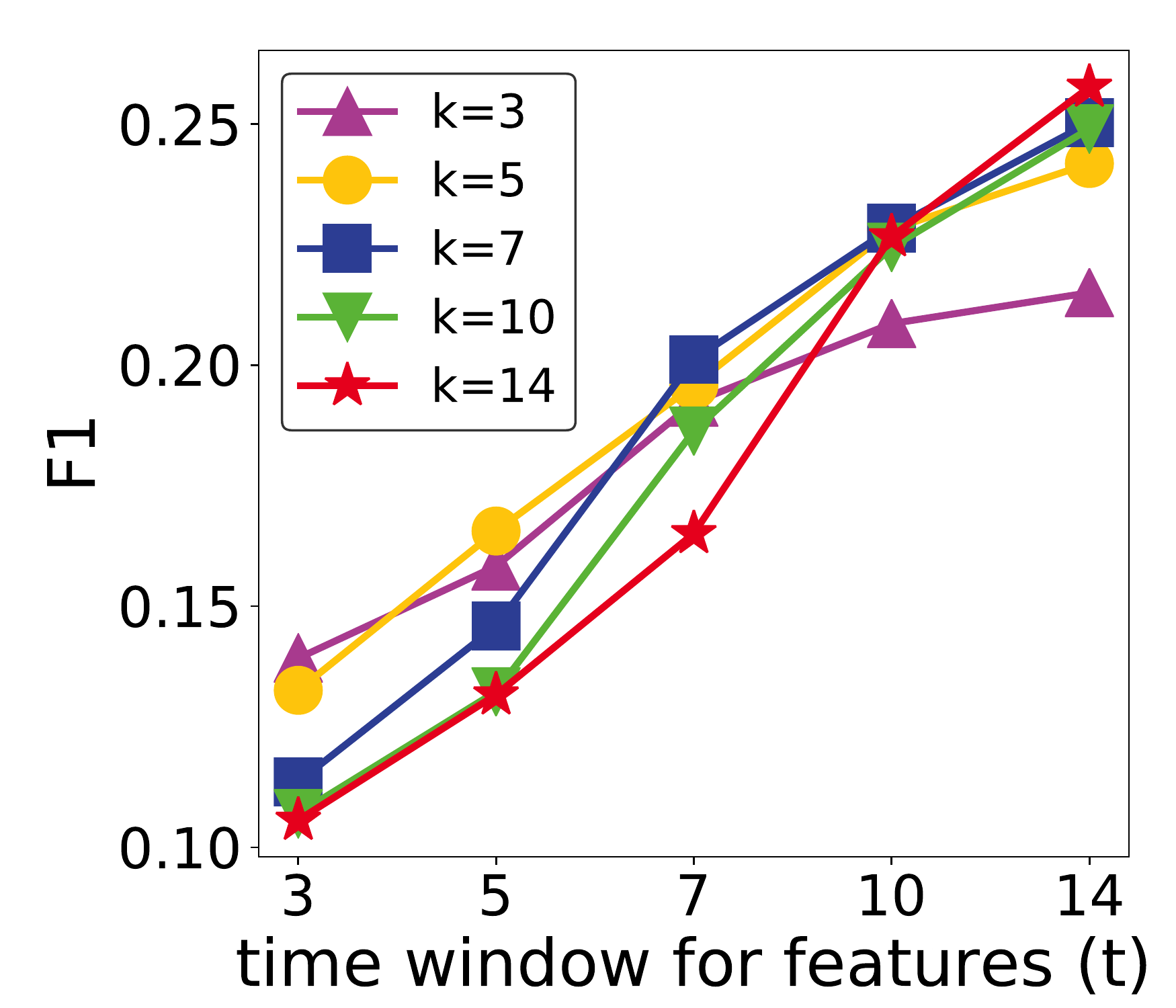, width=0.23\textwidth}
}
\caption{The left three figures present prediction performance in distinguishing \leaves from \stays by employing features extracted from the first $k$ days since new migrants moved to Shanghai. 
x-axis represents the number of days that we extract features from and train classifiers on, y-axis represents the evaluation metric.
Figure~\ref{fig:time} shows that the performance improvement mainly comes from improved feature quality.
x-axis represents the number of days ($t$) that we extract features from when testing, while different lines track the performance of the classifier trained using $k$ days.
The classifier with small $k$ (i.e., $\geq 5$) shows similar performance with $k=15$ as long as we use features from 14 days when testing.
}
\label{fig:prediction}
\end{figure*}

\subsection{Predicting Leaving Migrants}
\label{sec:exp:leave}
In the second task, we aim to separate \stays from \leaves, i.e., predict if a new migrant will leave Shanghai within the 3rd week.
We again  conduct 5-fold cross-validation and  use precision, recall, and F1-score for evaluation, with \leaves as the target class. 

\begin{table}[t]
	\begin{tabular}{lrrr}
		\toprule
		Classification method & Precision &   Recall   &  F1 \ \\ 
		\midrule
		Random forest & 0.1597 & 0.6659 & 0.2576 \\  
		Multilayer perceptron & 0.1329 & 0.5533 & 0.2140\\  
		Support vector machine & 0.1238 & 0.6815 & 0.2095 \\ 
		Logistic regression & 0.1006 & 0.7082 & 0.1762 \\ 
		\midrule
		random guess &0.0437 & 0.0426 & 0.0431 \\
		\bottomrule
	\end{tabular}
	\caption{Performance in distinguishing \leaves from \stays with different classifiers. Each classifier uses all features extracted from the first $k=14$ days. }
	\label{tb:classifier}
\end{table}

\begin{table}[t]
	\begin{tabular}{lrrr}
		\toprule
		Feature sets & Precision &  Recall &  F1 \ \\ 
		\midrule
		all features & 0.1597 & 0.6659 & 0.2576 \\ 
		ego network properties & 0.1347 & 0.6580 & 0.2234 \\
		housing price information & 0.1067 & 0.5978 & 0.1809 \\ 
		call behavior & 0.0984 & 0.5853 & 0.1683 \\ 		
		geographical information & 0.0863 & 0.5691 & 0.1498\\ 
		\bottomrule
	\end{tabular}
	\caption{Distinguishing \leaves from \stays using random forest with different feature sets extracted from the first $k=14$ days.}
		\label{tb:performance}
\end{table}


\vpara{Overall performance.}
We experiment with different classifiers including logistic regression, support vector machine, multilayer perceptron, and random forest.
Recall that we have 34K \stays and 1.5K \leaves in our dataset.
Random guessing would thus obtain an F1-score around 0.04.
Table~\ref{tb:classifier} shows that all the machine learning classifiers clearly outperform random guessing and random forest provides the best performance.
Random forest and MLP outperform others in terms of F1, suggesting that non-linearity is important for this classification task.

Overall, the prediction performance suggests that the proposed features are not only effective in distinguishing new migrants from locals, but also useful in predicting a migrant's decision to stay or leave.
However, the F1-score in this task is not as good as in the first task.
The performance drop suggests that as expected, predicting a migrant's leaving decision is much harder than distinguishing new migrants from locals. 

Table~\ref{tb:performance} also lists the performance of the classifier using a single feature set with random forest.
Again, ego network properties perform the best. 
Geographical patterns, however, perform worse in this task than in the task of distinguishing new migrants from locals. 
In comparison, housing price features achieve a better F1-score than geographical patterns, which suggests that knowing meta information such as housing price of a person's active areas is more useful than simply knowing the active areas for predicting migrants' early departure.



\vpara{Early detection of \leaves.} 
We next explore whether it is possible to detect \leaves sooner than two weeks.
If we can detect departure early on, we may be able to provide integration service.
To do that, we extract features based on a person's information from the first $k$ days.
Figure~\ref{fig:prediction} shows that precision and F1 score follow a very similar trend: 
knowing a person's behavior for longer after she moves to a new city (as $k$ increases) allows us to better predict her decision to leave or stay.
Since \leaves only take a very small fraction of our data,
recall is relatively stable over different $k$s, which is already around 0.6 when $k=3$, whereas improvement in precision is the main reason for overall performance improvement.
Even when observing only 3 days, the classifier can outperform random guessing. 


\vpara{Why does the performance improve?} 
To understand why the performance improves as we observe new migrants for longer, we propose a novel set of experiments.
We attempt to disentangle the improvement due to feature quality or classifier quality by replacing the features with future information when applying the classifier trained from only a small number of days.
Specifically, we first train a classifier with data in the first $k$ days, 
and then use features extracted from the first $t$ days to predict if the user will leave the city within the 3rd week after she arrives in  Shanghai. 
We vary different $k$ and $t$ to see how they influence the performance. 

Surprisingly, we see that classifiers trained with only the first 5 days' data perform as well as those trained using 14 days when testing with features extracted from the first 10 or 14 days in Figure~\ref{fig:time}. 
This result indicates that the classifier can be well trained using data from a small number of days and the performance improvement is mainly due to improved feature quality.
In other words, as new migrants stay longer, we have more reliable information regarding how she behaves, but even with less reliable information from the first 5 days, we can already know how different features relate to \leaves.

\begin{figure}[!t]
\centering
\epsfig{file=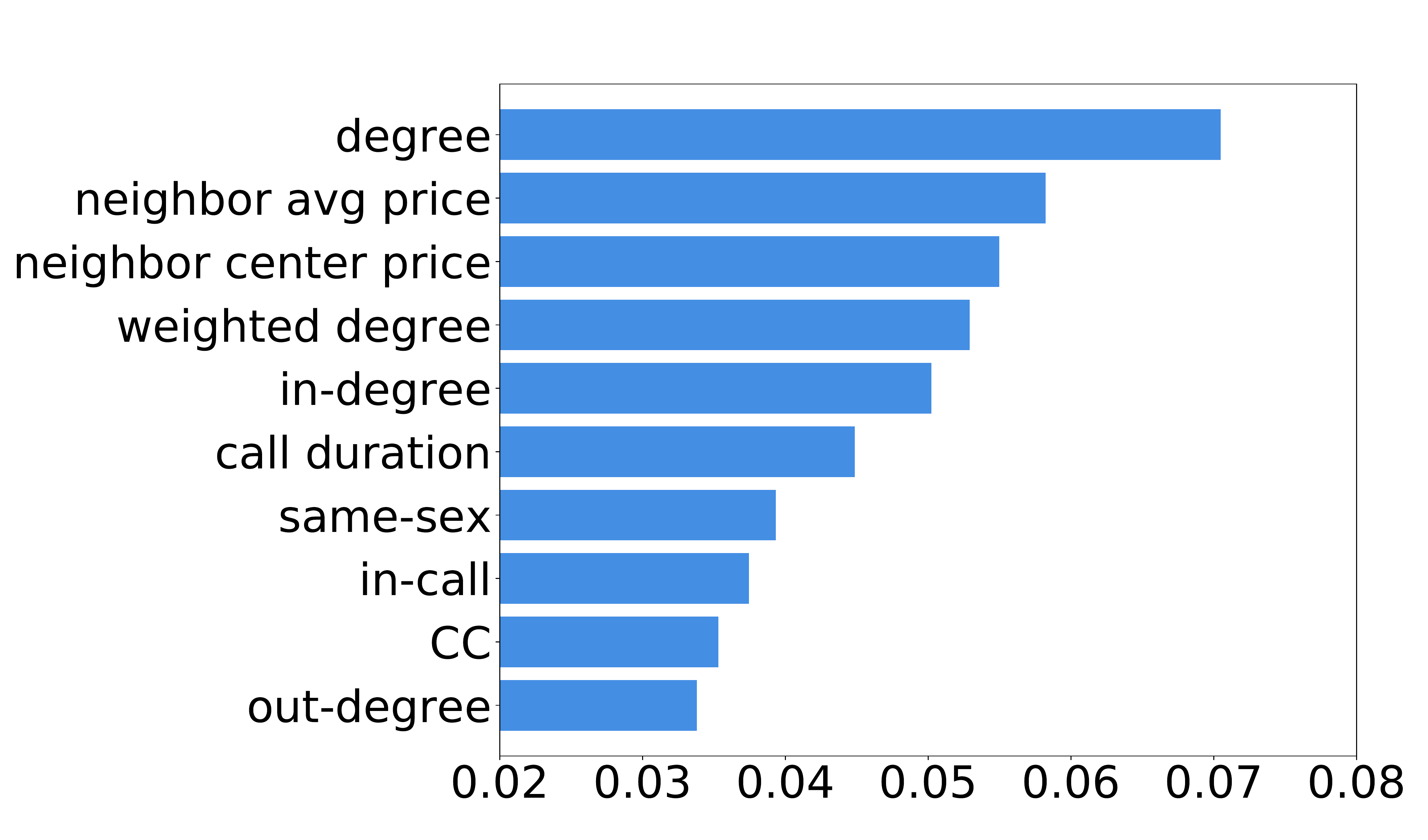, width=0.35\textwidth}
\caption{The most important 10 features. The x-axis denotes the relative Gini importance of features.}
\label{fig:top_feature}
\end{figure}

\vpara{Feature importance}
We finally discuss the important features in the learned random forests.
For each feature, we calculate its Gini importance (also known as mean decrease impurity) in the learned random forest model with the first 14 days' data. 
Figure~\ref{fig:top_feature} lists the most important 10 features. 
Degree (\#contacts) ranks the first, while in-calls (\#incoming calls), out-degree (\#contacts making outgoing calls), and CC (clustering coefficient) are both in top 10.
It, again, suggests that expanding more (diverse) connections is critical for new migrants and ego network properties are useful features. 
Two of the top 3 features are relevant to housing price, which is consistent with our finding in Section~\ref{sec:analysis} that living in an area with reasonable pricing is important. 

\section{Related Work}
\label{sec:related}

The urbanization process poses significant challenges for the society that require efforts from various disciplines. 
We summarize relevant studies in the following four aspects.

\para{Migrant integration.} 
Migrant integration is a well-recognized research question in many disciplines, including anthropology, economics, sociology, and urban planning \cite{castles2013age}.
Most relevant to our work is the study of urban migration
\cite{brown1970intra,schiller2009towards,Fischer:ToDwellAmongFriendsPersonalNetworks:1982,schiller2011locating,scholten2013agenda,brockerhoff1995child,glaeser2001cities,Goodburn:InternationalJournalOfEducationalDevelopment:2009,yang2018}. 
For example, our previous work explores different characteristics of locals, settled migrants, and new migrants~\cite{yang2018}. 
In addition to the effect of demographics (ethnic groups, rural vs. urban) on urban migrant integration, 
\citet{schiller2009towards} argue that the role of migrants in the cities depends on the rescaling of the cities themselves.
Government policy and agenda-setting also play an important role in the integration process \cite{scholten2013agenda}.
Beyond our scope, immigrants (migrants to a new country) and refugees (a subgroup of immigrants) have also received significant interests \cite{becker2011challenge,bean2003america,waters2005assessing,jacobsen2003dual,strang2010refugee}.

\para{Urban migrants in China.}
The unprecedented speed of development and the huge population in China have sparked
a battery of studies on urban migrants in China \cite{afridi2015social,Liu:HabitatInternational:2012,liang2001age,Chen:Cities:2015,wu2004sources,wu2007inequality,wu2004household,knight1999chinese,wang1999inside,ZHANG:ChinaEconomicReview:2003,yue2013role}.
There are at least three perspectives as suggested in \cite{knight1999chinese}: those of the migrants themselves, of the urban employers, and of the government.
Our work presents the perspective of migrants based on their telecommunication patterns.
It is worth noting that a central topic in public policy regarding migrants in China is the impact of the ``hukou'' system, a household registration system that limits the benefits and social welfare of migrants \cite{afridi2015social,wu2007inequality,wu2004household}.
Finally, although satisfying migrants' needs is among the challenges that \citet{Bai:NatureNews:2014} highlight in the Chinese government's urbanization strategy, little attention is paid to the social integration of migrants.

\para{Urban computing.}
Data-driven studies related to cities have gained importance recently and led to a new term, urban computing \cite{Quercia:2015:DLW:2736277.2741631,afridi2015social,DBLP:journals/corr/DredzeGRM16,Zheng:2014:UCC:2648782.2629592,jiang2013review,reades2007cellular,zheng2011urban,Hristova:2016:MUS:2872427.2883065}.
These studies combine heterogeneous data sources, including location data, social media activity data, mobile phone data and survey data, to propose metrics for city development and potentially guide urban policies.
For instance, 
\citet{zheng2011urban} employ GPS data from taxicabs to evaluate transportation system in Beijing;
\citet{de2016death} use mobile phone data to extract human activity and propose metrics to measure urban diversity.
Recently, Twitter has also been used as a tool for 
studies in understanding the global mobility patterns \cite{DBLP:journals/corr/DredzeGRM16}.

\para{Temporal social networks and online communities.}
Our work is also relevant to studies on the evolution of networks \cite{kossinets2006empirical,Jacobs:2015:ATU:2786451.2786477,Leskovec:2008:MES:1401890.1401948,leskovec2007graph,milo2004superfamilies,Paranjape:ProceedingsOfWsdm:2016,Viswanath:2009:EUI:1592665.1592675}.
Using data from online social media, these studies explore the connection between individual behavior and global network properties.
For instance, \citet{Viswanath:2009:EUI:1592665.1592675} find that links in the activity network tend to come and go rapidly over time, and the strength of ties exhibits a general decreasing trend 
as the social network link ages on Facebook.
\citet{Leskovec:2008:MES:1401890.1401948} develop a triangle-closing model to explain network evolution.
In addition, studies have investigated the process of new user integration in online communities \cite{lampe2005follow,McAuley:2013:ACM:2488388.2488466,Danescu-Niculescu-Mizil:2013:NCO:2488388.2488416,tan2015all,Backstrom:2006:GFL:1150402.1150412,Crandall:2008:FES:1401890.1401914}.
In particular, \citet{McAuley:2013:ACM:2488388.2488466} examine the process of how a new user becomes an expert on review websites.


\section{Conclusion}
\label{sec:conclude}
In this paper, we examine the integration process of migrants in the first weeks after moving to a new city, and the disintegration process for some migrants who left early. 
We use Shanghai as a case study and employ a one-month complete dataset of telecommunication metadata in Shanghai with 54 million users and 698 million call logs, plus a housing price dataset for 20K real estates in Shanghai. 
We examine four types of features extracted from people's mobile communication networks and geographical locations: 
ego network properties, call behaviors, geographical patterns, and housing price information.
Through our study, we find that some behavior patterns of \stays evolve towards \locals over time (e.g., contacting similar-aged people) as new migrants integrate into a new city.  
The communication networks that a new migrant develops is associated with her decision to stay or leave.
New migrants who manage to stay tend to develop diverse connections, move around the city in less expensive housing areas in the first weeks, compared to \leaves. 
Our proposed features are effective in both distinguishing new migrants from locals and distinguishing \leaves from \stays.
It is more challenging to predict \leaves than to predict new migrants in general.
The performance improves as we are able to observe new migrants' behavior for longer.
Intriguingly, when using the same features, the classifier trained from only the first few days is already as good as the classifier trained using full data.
This observation indicates that the performance improvement mainly lies in having better features.

Our work is limited by the data that we have access to.
We provide the large-scale characterization of the first weeks after migrants move to a new city, but the integration process takes much longer, maybe even a life.
Moreover, people's lives are much richer and more dynamic than what we are able to capture using telecommunication metadata.
Job situations, health records, and daily interactions can all potentially offer a more in-depth understanding of the integration process.
Last but not least, although China Telecom is a major service provider and Shanghai is an important global city, the selection bias in our data may limit the generalizability of our findings.

As urbanization is happening at an unprecedented rate and data collection becomes ubiquitous in smart cities, there are tremendous opportunities for data-driven approaches to understanding and improving migrant integration.
For instance, it would be useful to identify what difficulty a particular migrant have in fitting into the city and provide timely and useful support.

\small
\vpara{Acknowledgements.}
The work is supported by 
the Fundamental Research Funds for the Central Universities, 
NSFC (61702447, 
U1611461, 61625107), 
and a research funding from HIKVISION Inc. 
\normalsize

\appendix
\vspace{-0.1in}
\section{Appendix} 
In Table~\ref{tb:feature}, we list features that we explore in Section~\ref{sec:analysis} and use for the classification tasks in Section~\ref{sec:exp}. 
We omit simple demographics features based on the user's personal attribute such as age and gender in the table for space reasons. 
We view all directed edges as undirected except in measuring reciprocal calls.
For demographics related features, we only include users for whom we have the corresponding information.
\begin{table}[H]
 \centering 
 \small
 \begin{tabular}{p{0.12\textwidth}p{0.34\textwidth}}
 \toprule
 Feature &  Description \\
 \midrule
\multicolumn{2}{c}{\textbf{Ego networks of user $v$ in $G_t$}
} \\
 similar-age & The fraction of $v$'s contacts that are at similar ages with $v$ ($\pm 5$ years). \\
 same-sex & The fraction of $v$'s contacts with the same sex with $v$. \\
 local & The fraction of $v$'s contacts born in Shanghai. \\
 townsman & The fraction of $v$'s contacts born in the same province with $v$ but not in Shanghai. This feature is always 0 for locals, so it is not included in prediction experiments in Section 4.1.\\
 degree & The number of $v$'s unique contacts. \\ 
 in(out)-degree & The number of $v$'s unique contacts having been called by $v$ (called $v$)\\
 neighbor degree & The average degree of $v$'s contacts. \\
 CC  & Clustering coefficient of $v$'s ego-network, $\frac{|\{(s, t)| (s, t) \in E_t\}|}{d_v(d_{v}-1)}$, where $s$ and $t$ are $v$'s contacts, and $d_{v}$ is $v$'s degree. \\
 \midrule 
 \multicolumn{2}{c}{\textbf{Call behavior of user $v$ in $G_t$}} \\
 in(out)-call & The number of incoming (outgoing) calls. \\
 out-call - in-call & The difference between the number of outgoing calls and incomming calls.  \\
 (local) call duration & $v$'s average call duration (with locals). \\
 (local) duration variance & The variance of $v$'s call duration (with locals).\\
 province diversity & Entropy of the distribution of birth provinces 
 among $v$'s contacts, defined as $-\sum_i p_i \log_2 p_i$, where $p_i$ is the probability that a contact of $v$ was born in province $i$.\\
  reciprocal call & The probability that $v$'s contacts also call $v$.
  \\
 communication diversity & Shannon entropy of the distribution of the number of calls to their contacts, defined as $\frac{-\sum_{j}p_{ij}\log(p_{ij})}{log(k_i)}$,
 where $k_i$ is the out-degree,  
 $p_{ij}=\frac{n_{ij}}{\sum_{l}n_{il}}$, $n_{ij}$ is the number of calls user $v_i$ makes to user $v_j$. \\
 \midrule 
 \multicolumn{2}{c}{\textbf{Geographical features of $v$ at time $t$}} \\
 center & The latitude and longitude of a user $v$'s center of mass $l_{\operatorname{CM}}$,
 $l_{\operatorname{CM}}=\frac{1}{|L_v^t|}\sum_{l \in L_v^t}l$.
 \\
 workplace center & The center of user $v$ during 9:00am to 16:00pm\\ 
home center & The center of user $v$ during 20:00pm to 7:00am\\ 
average radius & The average distance of $v$ from her center of mass, i.e., $\frac{1}{|L_v^t|}\sum_{l \in L_v^t} |l - l_{\operatorname{CM}}|$.\\
max radius & The maximal distance of $v$ from her center of mass, i.e., $\max_{l \in L_v^t} |l - l_{\operatorname{CM}}|$. \\
 moving distance & The total distance that $v$ moves, $\sum_i |l_i - l_{i-1}|$. \\
 average distance & The average distance that $v$ moves, $\frac{1}{|L_v^t|}\sum_i |l_i - l_{i-1}|$. \\
 home distance & The distance between $v$'s workplace and home.\\
 \midrule
 \multicolumn{2}{c}{\textbf{Housing price features of user $v$}} \\
 average price & The average housing price of $v$'s active areas. \\
 center price & The housing price of $v$'s center of mass. \\
 neighbor avg(center) price & The average value of the average(center) price  of $v$'s contacts. \\
 workplace avg(center) price & The average(center) price of user $v$ during 9:00am to 16:00pm. \\
 home avg(center) price & The average(center) price of user $v$ during 20:00pm to 7:00am. \\
 \bottomrule 
 \end{tabular}
 \caption{
 List of features. 
 }
 \label{tb:feature} 
 \end{table}

\clearpage
\bibliographystyle{ACM-Reference-Format}
\bibliography{reference}

\normalsize

\end{document}